\documentclass[11pt]{article}
\usepackage[nosort]{cite}
\usepackage{dsfont}
\usepackage{amsfonts}
\usepackage{amssymb}
\usepackage{amsmath}
\usepackage{graphicx}
\usepackage{slashed}
\usepackage{setspace}
\usepackage{multirow}

\newcommand{\myhline}{~\\[-2mm]\hline\\[-2mm]}

\newcommand{\barW}{{\overline{W\!}\,}}
\newcommand{\ldotss}{\!.\,.\,}

\usepackage{color}

\newcommand{\cn}{{\cal N}}

\renewcommand{\baselinestretch}{1.05}
\def\pslash{\slash \negthinspace \negthinspace \negthinspace \negthinspace  p}

\newcommand{\reef}[1]{(\ref{#1})}

\renewcommand{\baselinestretch}{1.1}
\newcommand{\dash}{\text{-}}

\newcommand{\be}{\begin{equation}}
\newcommand{\ee}{\end{equation}}

\def\be{\begin{equation}}
\def\ee{\end{equation}}
\def\bea{\begin{eqnarray}}
\def\eea{\end{eqnarray}}
\def\ba{\begin{array}}
\def\ea{\end{array}}
\def\bd{\begin{displaymath}}
\def\ed{\end{displaymath}}

\def\eg{{\it e.g.~}}
\def\ie{{\it i.e.~}}
\def\Tr{{\rm Tr}}

\def\>{\rangle} 
\def\<{\langle} 
\def\Dsl{D \hskip-.6em \raise1pt\hbox{$ / $ } }
\def\to{\rightarrow}

\def\lab{\label}

\newcommand{\eps}{\epsilon}

\def\tQ{\tilde{Q}}

\begin{document}

\setstretch{1.05}

\begin{titlepage}

\begin{flushright}
MCTP-11-38\\
MIT-CTP-4322  \\
PUPT-2397
\end{flushright}
\vspace{.5cm}

\begin{center}
\bf \Large
Integrands for QCD rational terms and $\cn=4$ SYM\\[1ex] from massive CSW rules
\end{center}
\vskip1cm
\begin{center}
{\bf Henriette Elvang${}^{a}$,
Daniel Z.~Freedman${}^{b,c}$, Michael Kiermaier$^{d}$} \\
\vspace{0.7cm}
{${}^{a}${\it Michigan Center for Theoretical Physics, Randall Laboratory of Physics}\\
{\it University of Michigan, Ann Arbor, MI 48109, USA}}\\[2mm]
{${}^{b}${\it Department of Mathematics, }{${}^{c}${\it Center for Theoretical Physics,}}\\
{\it Massachusetts Institute of Technology, Cambridge, MA 02139, USA}}\\[2mm]
{${}^{d}${\it Joseph Henry Laboratories, Princeton University, Princeton, NJ 08544, USA}}\\[3mm]
{\small \tt  elvang@umich.edu,
 dzf@math.mit.edu, mkiermai@princeton.edu}
\end{center}
\vskip2cm

\begin{abstract}
\vskip.2cm
\setstretch{1.05}
We use massive  CSW rules to derive explicit compact expressions for  integrands of rational terms in QCD with any number of external legs. Specifically, we present all-$n$ integrands for the one-loop all-plus and one-minus gluon
amplitudes
in QCD.
We extract the finite part of spurious external-bubble contributions systematically; this is crucial for the application of integrand-level CSW rules in theories without supersymmetry. Our approach yields integrands that are independent of the choice of CSW reference spinor even before integration.\\[-1ex]

Furthermore, we  present a
recursive
derivation of the recently proposed  massive CSW-style vertex expansion for massive tree amplitudes and loop integrands on the Coulomb-branch of $\cn=4$ SYM. The derivation requires a careful study of boundary terms in all-line shift recursion relations, and provides a rigorous (albeit indirect) proof of the recently proposed  construction of massive amplitudes from soft-limits of massless
on-shell amplitudes. We show that the massive vertex expansion manifestly preserves all holomorphic and half of the anti-holomorphic supercharges, diagram-by-diagram, even off-shell.
\end{abstract}

\end{titlepage}

\vspace{4mm}
\setstretch{0.3}
\setcounter{tocdepth}{2}
\tableofcontents
\setstretch{1.05}

\setcounter{equation}{0}
\section{Introduction}
The CSW expansion~\cite{Cachazo:2004by}, or MHV vertex expansion, has proven to be a valuable tool in the study of massless amplitudes in gauge theory~\cite{Brandhuber:2004yw,Risager:2005vk,Elvang:2008na,Elvang:2008vz,Bullimore:2010pj,Bullimore:2010dz,Mason:2010yk} and beyond~\cite{Cohen:2010mi}.  In this paper, we use massive  CSW-style vertex expansions to study amplitudes in
 massless
 QCD and  in $\cn=4$ SYM on the Coulomb branch.
The
 massive CSW expansions
in these two theories are related even though masses appear for  very different reasons in these two cases. Massive particles are naturally part of the spectrum of $\cn=4$ SYM on the Coulomb-branch, where the gauge and $R$-symmetry groups are spontaneously broken. On the other hand, in massless QCD one-loop amplitudes, particles running in the loop  effectively acquire masses because in dimensional regularization the  $(D-4)$-dimensional components of the $D$-dimensional loop momentum
 can be encoded in a mass-term~\cite{Badger:2008cm}.
Thus we use massive vertex rules for amplitudes and loop-integrands in both theories. In fact, the non-supersymmetric rules for the  QCD integrand turn out to be a simple special case of the fully supersymmetric $\cn=4$ rules.

The {\em massless} CSW expansion is well-known  to produce the correct QCD gluon amplitudes at tree level~\cite{Risager:2005vk}; however, it fails to produce the full loop integrand when applied naively to non-supersymmetric theories. At one loop, for example, the massless CSW expansion of QCD loop integrands misses the crucial {\em rational terms}, which are not cut-constructible in $4$ dimensions~\cite{Ettle:2007qc,Brandhuber:2007vm}. This failure seems to be closely related to the breakdown of loop-level recursion relations for integrands in non-supersymmetric theories (caused by infinite forward-limit contributions~\cite{CaronHuot:2010zt,ArkaniHamed:2010kv}). One way to construct rational terms is to  apply dimensional regularization, effectively giving a mass $\mu$ to internal lines in one-loop amplitudes~\cite{Badger:2008cm}.
The mass is then integrated over with an appropriate measure. As we will review below, it is sufficient to consider a charged massive scalar running in the loop to compute rational terms.    A  massive vertex expansion similar to CSW  was developed for such diagrams in~\cite{Boels:2007pj}. (See~\cite{NigelGlover:2008ur} for applications at the 4- and 5-point level.) Using this  massive vertex expansion and the techniques developed in~\cite{Kiermaier:2011cr}, we derive an extremely compact all-$n$ expression for the integrand ${\cal I}^{++\cdots+}$ of the (purely rational) all-plus amplitude in QCD.  We also present a similarly compact all-$n$ BCFW-like representation of the same integrand, which is manifestly free of spurious poles.  Readers may wish to peek at \reef{allplusCSW}
and \reef{allplus2} for explicit expressions of our CSW and BCFW all-plus integrands.

We then turn to the (also purely rational) one-minus integrand ${\cal I}^{-+\cdots+}$. Here the  massive vertex expansion cannot be applied naively; divergent ``external-bubble diagrams'', which integrate to zero in dimensional regularization in conventional Feynman-gauge diagram computations, can no longer be ignored.   Indeed, these diagrams contain spurious poles and do not integrate to zero. We use  a systematic approach inspired by unitarity methods~\cite{Bern:1994zx,Bern:1994cg} to construct finite external-bubble-like ``counterterms'' that supplement the naive  massive vertex rules. These counterterms ensure that the one-minus integrand is free of spurious poles.  Correct factorization properties are also maintained. This leads to a rather compact all-$n$ expression for the integrand of the
 one-minus QCD amplitude;  see \reef{oneminus}.
 The counterterms have features that indicate a possible interpretation as the finite parts of  divergent ``external-bubble diagrams"; it would be interesting to clarify this connection.

  Why should we bother determining integrands for rational terms in QCD?  After all, explicit expressions are known for the integrated results to leading order in $\epsilon$~\cite{Bern:1993qk,Mahlon:1993si,Bern:1995db,Bern:2005hs,Bern:2005ji,Brandhuber:2007up,Cohen:2010mi}. The motivation for our analysis is two-fold: first of all, our analysis gives the full integrand,
  which factorizes correctly into tree amplitudes and is
  valid to all orders in $\epsilon$.\footnote{
   The integrand of the all-plus QCD amplitude has been conjectured to satisfy a curious ``dimension-shifting'' relation to the integrand of one-loop MHV amplitudes in $\cn=4$ SYM~\cite{Bern:1996ja}. This relation is not manifest in our approach.
   }
   This all-order integrand could be useful, for example, as input to determine higher-loop integrands in QCD.
  Secondly, note that the recently found recursion relations \cite{ArkaniHamed:2010kv,Boels:2010nw}  for planar loop-integrands require a well-defined forward limit; this can be achieved in supersymmetric theories~\cite{CaronHuot:2010zt}, but fails in non-supersymmetric cases, for example for the one-minus amplitude  QCD.    Thus we regard our result for the one-minus integrands as
   a
  non-trivial  step towards applying  recursive techniques to integrands in non-supersymmetric theories
  (see also~\cite{Britto:2011cr}).
  Our integrands can therefore serve as
   valuable ``data points''
  for loop-level recursive methods in QCD. A challenge that remains is the direct integration of  the all-$n$ integrands we construct. Both standard integral reduction and Badger's method~\cite{Badger:2008cm} are viable approaches. However, our integrands (and generalizations thereof) would be  more  useful  if terms with spurious singularities could be integrated directly.

In the second part of the paper, we study $\cn=4$ SYM in its spontaneously-broken phase, the Coulomb-branch. Coulomb-branch amplitudes have recently been studied (i) as an infrared regulator~\cite{Alday:2009zm, Sever:2009aa, Henn:2010bk,Henn:2010ir,Henn:2011xk} for massless planar integrands in $\cn=4$~\cite{Bern:1997nh,Bern:2005iz,Bern:2006ew,Bern:2007ct}, (ii) because they arise in the dimensional reduction of the massless  maximally supersymmetric $6$-dimensional $(1,1)$ theory~\cite{Bern:2010qa,Brandhuber:2010mm,Dennen:2010dh,Hatsuda:2008pm,Elvang:2011fx,Huang:2011um}, and (iii), in their own right, as the ``simplest'' massive field theory in 4 dimensions~\cite{Schabinger:2008ah, Boels:2010mj,Boels:2011zz,Craig:2011ws,Kiermaier:2011cr}.
 A direct construction of massive Coulomb-branch tree amplitudes and loop integrands from massless on-shell amplitudes was proposed in~\cite{Craig:2011ws,Kiermaier:2011cr}. It was shown in~\cite{Kiermaier:2011cr} that this construction implies a certain massive vertex expansion for Coulomb-branch amplitudes, which we review
 in section \ref{secCoulombReview}.  In this paper, we  derive this  expansion for tree amplitudes and loop integrands from recursion relations.
 Specifically, we use recursion relations based on an {\em anti-holomorphic} all-line shift
 $|i] \to |i] + z\, b_i |q]$~\cite{Elvang:2008vz,Bullimore:2010dz}
 (see also~\cite{Kiermaier:2009yu,Cohen:2010mi}) to construct the diagrammatic expansion up to tree-level boundary terms. We then recursively construct the missing boundary terms from a {\em holomorphic}
 all-line shift
 $|i\> \to |i\> + w\, \tilde{b}_i |q\>$.
 In particular, our derivation provides a (somewhat indirect) proof of the soft-limit construction of Coulomb-branch amplitudes proposed in~\cite{Craig:2011ws,Kiermaier:2011cr}.

We also study the supersymmetry properties of the massive vertex expansion on the Coulomb branch of $\cn=4$ SYM. We find that it manifestly preserves all anti-holomorphic and half of the holomorphic supercharges of the Coulomb-branch SUSY algebra, diagram-by-diagram, even off-shell. This matches the SUSY properties of the massless CSW expansion.
As a consequence, any diagram with a self-energy-type subdiagram vanishes.
This property greatly facilitates our loop-level derivation of the expansion, and reduces the number of diagrams that appear in the expansion of loop integrands.
The massive vertex expansion procedure is well-suited for automatization in computer-codes, so this could be used to compute actual loop integrands and amplitudes on the Coulomb-branch of $\cn=4$ SYM to all orders in the mass.

\setcounter{equation}{0}
\section{All-$n$ integrands for rational terms in QCD}\label{secRational}
\setcounter{equation}{0}
\subsection{Review: rational terms from massive scalar amplitudes}\label{secratscal}
It is  well known
that 1-loop
gluon
amplitudes in pure YM can be decomposed into a sum of $\cn=4$, $\cn=1$, and  scalar ($\cn=0$) amplitudes in the following way:
\begin{equation} \label{decomp}
    {\cal A}_n^{\text{pure YM}}~=~{\cal A}_n^{\cn=4}-4\,{\cal A}_n^{\cn=1}+{\cal A}_n^{\text{scalar}}\,  ,
\end{equation}
where the superscript indicates what runs in the loop.
The ``scalar"-label indices a  complex scalar canonically coupled to the gluons.

Throughout this section, we focus on all-plus and one-minus color-ordered gluon amplitudes,
${\cal A}_n^{++\cdots+}$ and ${\cal A}_n^{-+\cdots+}$.
These vanish in supersymmetric theories and by \reef{decomp}
can therefore be computed directly from the third term ${\cal A}_n^{\text{scalar}}$ alone.
In pure Yang-Mills theory, only gluons run in the loop; with the prescription \reef{decomp} the internal gluon is replaced by the complex scalar.

In massless QCD,
$n_f$
flavors of
massless
``quarks'' in the fundamental representation
circulate in the loop
in addition to the gluons.
Hence the massless QCD gluon amplitude
 is related the the pure YM
gluon
 amplitude by a factor of
$N_p\equiv 1-n_f/N_c$. We can thus perform the computation in pure YM and obtain the QCD result simply by multiplying the result by $N_p$:
\begin{equation}
    {\cal A}_n^{\text{QCD}}\,=~N_p\,{\cal A}_n^{\text{pure YM}}\,=~N_p\,{\cal A}_n^{\text{scalar}} \qquad
    \text{(for all-plus and one-minus
    gluon
    amplitudes)}\,.
\end{equation}

All-plus and one-minus gluon amplitudes do not have any cut-constructible contributions in 4 dimensions,  because the product of tree amplitudes   in the cut loop integrand vanishes.  To compute
${\cal A}_n^{\text{scalar}}$,  one uses dimensional continuation to $D=4-2\epsilon$ dimensions.
The ($D-4$)-dimensional  components $\mu$ of the loop momentum
enter as  an effective 4-dimensional  mass of the scalar field; $\mu$ is then integrated over with an appropriate measure as part of the $D$-dimensional loop-momentum integration:
\begin{equation}
    {\cal A}_n^{\text{scalar}}~=~\int \frac{d^D\ell}{(2\pi)^D}~
    {\cal I}_n^{\text{scalar}}~=~
    \int \frac{d^4\ell}{(2\pi)^4}\int\frac{d^{-2\epsilon}\mu}{(2\pi)^{-2\epsilon}}
    ~
    {\cal I}_n^{\text{massive scalar}}\,.
\end{equation}
To compute all-plus and one-minus gluon amplitudes it thus suffices to determine the 1-loop integrand ${\cal I}_n^{\text{massive scalar}}$, which describes $n$ gluons interacting with a
massive scalar running in the loop. The massive CSW-style vertex expansion for gluons interacting with a massive scalar introduced in~\cite{Boels:2007pj} will be used in our computation of ${\cal I}_n^{\text{massive scalar}}$. We review this  massive vertex expansion approach now.

\subsection{Review: the CSW expansion with a massive scalar}
\label{s:reviewCSW}

Scattering amplitudes for gluons interacting with a charged massive scalar can be  computed conveniently from the  massive CSW rules given in~\cite{Boels:2007pj}. These rules can also be understood as a special case of the massive   CSW expansion on the Coulomb branch of $\cn=4$ SYM~\cite{Kiermaier:2011cr}, which we will derive in section~\ref{secCoulomb}. We emphasize that all momenta appearing in the CSW rules are
{\em strictly 4-dimensional}.
When we apply the CSW expansion to the dimensionally-regulated QCD amplitudes, the $(D-4)$-dimensional components of the momenta
arise only through the effective mass $\mu$ of the $4$-dimensional scalar particles. For the CSW diagrams of QCD gluon amplitudes, the massive particles only appear as internal lines, so we can use
the
conventional 4-dimensional massless spinor helicity formalism $p_i^{\dot\alpha \alpha}~=~|i\>^{\dot\alpha}[i|^{\alpha}$ for all external momenta (they are null!) and simply apply the usual CSW prescription
\begin{equation}\label{CSW}
    |P\>~\equiv~P|q]\,
\end{equation}
for the internal lines. This rule is used for any internal line
 momentum $P$ in the CSW diagrams, whether it is
massive or massless. The reference spinor $|q]$ appearing in this assignment can be chosen arbitrarily, but consistently for all
 internal
lines.
The sum of all contributing CSW diagrams  must be independent of $|q]$.

We can now state the CSW rules. Diagrams are built from vertices and scalar propagators. The scalar propagators are massless for gluon internal lines, and massive for scalar internal lines:
\begin{equation}
        \parbox[c]{1.8cm}{\includegraphics[width=1.6cm]{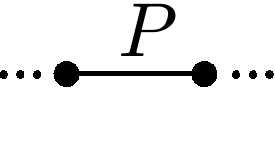}}~=~\frac{1}{P^2}\,,\qquad\qquad
        \parbox[c]{1.8cm}{\includegraphics[width=1.6cm]{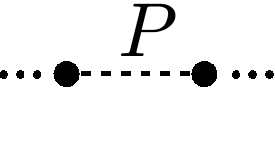}}~=~\frac{1}{P^2+\mu^2}\,.
        \end{equation}

There are 3 types of vertices in the massive CSW expansion:
\begin{itemize}
\item \emph{Gluon MHV vertex:}
This vertex has two negative-helicity gluons and arbitrarily many positive-helicity gluons, and is given by the familiar Parke-Taylor expression~\cite{Parke:1986gb}:
\begin{equation}
    \parbox[c]{2.2cm}{\includegraphics[width=1.8cm]{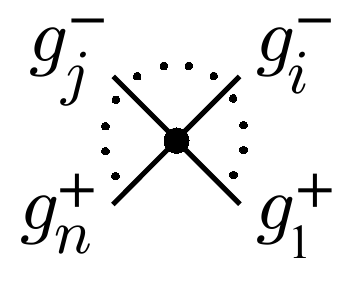}}~=~\frac{\<ij\>^4}{\<12\>\<23\>\cdots\<n1\>}\,.
\end{equation}
\item \emph{Scalar-gluon MHV vertex:}
This vertex couples a pair of conjugate scalars to one negative-helicity gluon and arbitrarily many positive-helicity gluons:
\begin{equation}
      \parbox[c]{2.2cm}{\includegraphics[width=1.8cm]{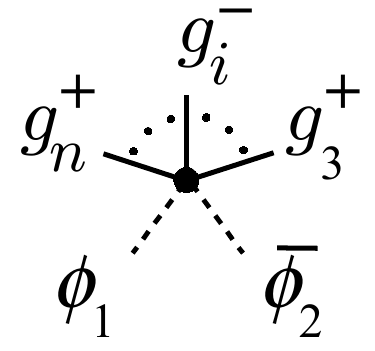}} ~=~\frac{\<1i\>^2\<2i\>^2}{\<12\>\<23\>\cdots\<n1\>}\,.
\end{equation}
The angle spinors $|1\>$ and $|2\>$ associated with the scalars lines are defined via the CSW prescription~(\ref{CSW}).
\item  \emph{Scalar-gluon ultra-helicity-violating (UHV) vertex:}
This couples a pair of conjugate scalars to arbitrarily many positive-helicity gluons:
\begin{equation}\label{UHVsc}
\parbox[c]{2.2cm}{\includegraphics[width=1.8cm]{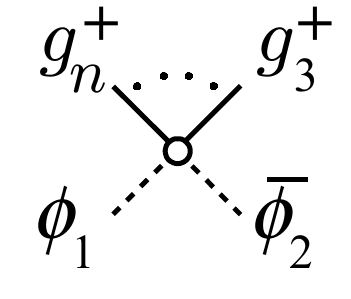}}   ~=~\frac{\mu^2\<12\>}{\<23\>\cdots\<n1\>}\,.
\end{equation}
The UHV vertex contains an explicit factor of $\mu^2$ and therefore vanishes in the massless limit $\mu\to0$. Interpreting $\mu$ as the $(D-4)$-dimensional component of the loop momentum, it is obvious that  diagrams with UHV vertices cannot be cut-constructible in 4 dimensions.
\end{itemize}

These  massive CSW rules can be used to compute tree-level amplitudes for scalar-gluon interactions. At the level of the loop integrand their application is  more subtle, but some progress was made in~\cite{NigelGlover:2008ur} at the 4- and 5-point level using a single-cut construction. In~\cite{NigelGlover:2008ur}, the reference spinor $|q]$ was  chosen in a very particular way to argue that certain (divergent) diagrams in the loop-integrand expansion integrate to zero and can thus be dropped. In the current work, we will keep the reference spinor $|q]$ arbitrary at all times, because $|q]$-independence can then be used as a tool to verify the absence of spurious poles.

\subsection{The all-plus integrand}\label{secallplus}
As a first application of the CSW rules to loop integrands, let us compute the
all-plus 1-loop integrand in QCD  for arbitrary $n$. As explained in section
 \ref{secratscal},
this amplitude can be computed from the contribution of  a massive
scalar running in the loop.
 The CSW-type diagrams needed are those involving only vertices with positive-helicity gluons as external states and massive scalars as internal lines.
 The diagrams are thus built from the
 UHV vertices~(\ref{UHVsc}) only.
 Each vertex must be a least cubic, so an $n$-point amplitude will consist of the sum of all diagrams with $k=2,\dots,n$ vertices.
 We use \emph{tadpoles} to denote diagrams with a single vertex and a closed scalar loop. Tadpole diagrams with  a UHV vertex are zero, because the numerator factor $\<12\>$ in~(\ref{UHVsc}) vanishes for $p_1=-p_2=\ell$.

To ensure that the loop-momentum $\ell$ is consistent between diagrams, we define $\ell$ as the momentum that flows between lines $1$ and $n$; clearly this is well-defined since the amplitude is color-ordered. We define
\bea
  \ell_i = \ell + \sum_{j=1}^i p_j\,,\qquad\text{with }~~\ell\equiv\ell_n
\eea
as convenient loop-momentum labels to be used in individual diagrams.

The sum over diagrams that contribute to the all-plus integrand takes the schematic form
\begin{equation}\label{allplusDiag}
 I_{\rm CSW}^{++\cdots+}~=~2N_p~\sum~~\parbox[c]{2cm}{\includegraphics[width=2cm]{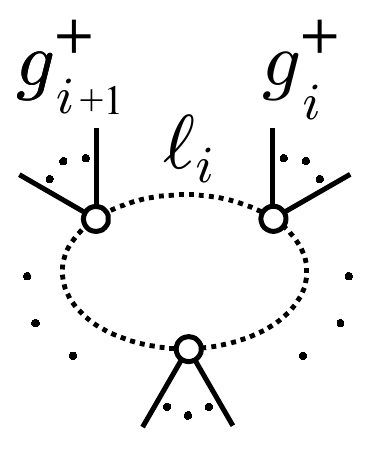}}
 \end{equation}
 Here, the factor of $2$ accounts for
two charged states of the scalar,
 and the factor $N_p$ converts the pure YM integrand into a QCD integrand, as explained above.
To illustrate the method, let us give
an example of the value of one diagram that contributes to the 4-point
all-plus integrand:
\begin{equation}
\begin{split}
  \parbox[c]{3.2cm}{\includegraphics[width=3.2cm]{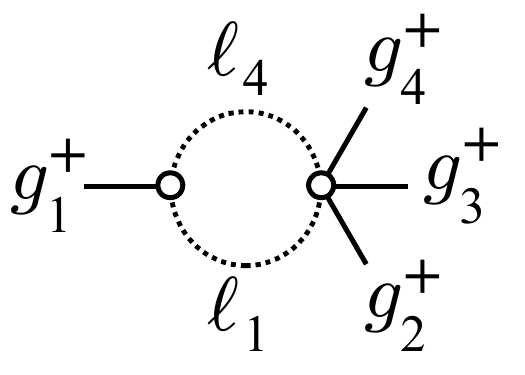}} ~=~~
  \frac{\mu^2 \< \ell_1 \ell_4\>}{\< \ell_4 1\>\<1 \ell_1\>}\times
  \frac{1}{(\ell_1^2 +\mu^2)(\ell_4^2 +\mu^2)}\times
  \frac{\mu^2 \< \ell_4 \ell_1\>}{\< \ell_1 2\>\<23\>\<34\>\<4 \ell_4\>  }
\end{split}
\end{equation}
with the CSW-prescription $|\ell_i\> \equiv \ell_i|q]$ understood.

One must add all possible diagrams of the type displayed in
\reef{allplusDiag}.
Their sum can actually be written in a very compact way. To see this, we first remind the reader about the tree-level CSW amplitude computation of
\cite{Kiermaier:2011cr}
in which a similar simplification occurred in the sum over all diagrams.
Consider the tree-level amplitude
\begin{equation}\label{WWn1}
\bigl\< \phi_1 \bar\phi_2\, g_3^+\ldots g_n^+\bigl\>_\text{tree}
~=~~
   \parbox[c]{2.5cm}{\includegraphics[width=2.5cm]{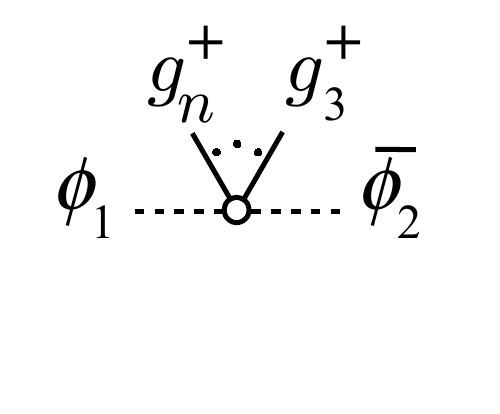}}
~~+~~\sum_{i=3}^{n-1}~\,\parbox[c]{3.75cm}{\includegraphics[width=3.75cm]{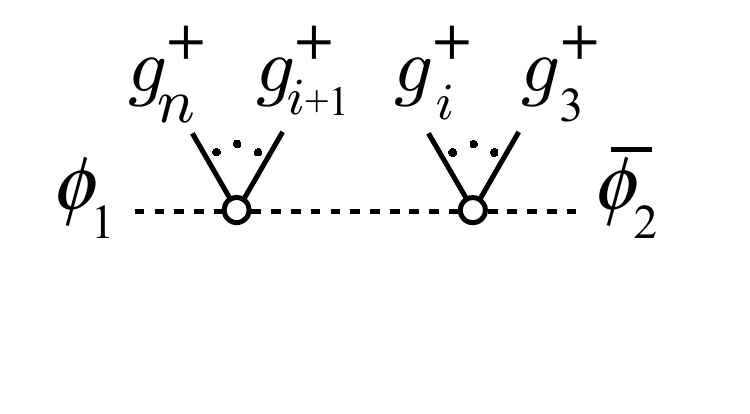}}
    +~\ldots\,,
\end{equation}
whose CSW-type expansion is illustrated on the right-hand side. The ``+\dots" stands for sums of CSW diagrams with $3,4,\dots,n$ blobs.
It was shown in \cite{Kiermaier:2011cr} that the full set of
diagrams in \reef{WWn1}
can be summed to the compact expression\footnote{The amplitude that was actually computed in~\cite{Kiermaier:2011cr} involved a pair of massive $W$-bosons and is trivially related to the given scalar amplitude by supersymmetry.}
\begin{equation}\label{WWalln1}
    \bigl\< \phi_1 \bar \phi_2\, g_3^+\ldots g_n^+\bigl\>_\text{tree} ~=~-
    \frac{\mu^2}{\<2 3\>\<34\>\cdots \<n1\>}\times \big\<2\big|\prod_{j=3}^{n-1}\biggl[1-\frac{\mu^2|P_J\>\<j,j\!+\!1\>\<P_J|}{(P_J^2+\mu^2)\<P_J,j\>\<j\!+\!1,P_J\>}\biggr]
    \big|1\big\>\,,~~
\end{equation}
with $P_J=P_{2\dots j}$\,.
 Here, the angle spinors $|1\>$ and $|2\>$ associated with external massive scalars are given by the CSW prescription,~(\ref{CSW}).
 Then note that the diagrams
\reef{allplusDiag}
 of the all-plus integrand are obtained by simply tying the massive scalar line of the above tree-amplitude \reef{WWn1} into a loop. Thus we simply need to trace the result \reef{WWalln1}  over the two-dimensional spinor space and relabel lines to find the all-$n$ expression for
the all-plus integrand! The result is
\begin{equation} \label{allplusCSW}
\boxed{~
\begin{split}
  I_{\rm CSW}^{++\cdots+}(1,\ldots,n)~=~\frac{2N_p}{\<12\>\cdots\<n1\>}\times
  \Tr'\prod_{j=1}^{n}\biggl[1-\frac{\mu^2|\ell_j\>\<j,j\!+\!1\>\<\ell_j|}{(\ell_j^2+\mu^2)\<\ell_j,j\>\<j\!+\!1,\ell_j\>}\biggr]\,,
\end{split}
~}
\end{equation}
where we defined
\begin{equation}
    \Tr'X~\equiv~\Tr\,X-\Tr\,1\,,
\end{equation}
to subtract the $\Tr\, 1=2 $ term in the trace~(\ref{allplusCSW}), because it does not correspond to any CSW diagram.
The integrand~(\ref{allplusCSW}) correctly factorizes into the tree amplitude~(\ref{WWalln1}) on the ``single cut'' of any  loop propagator $1/(\ell_i^2+\mu^2)$. As a further consistency check on the loop integrand, we have verified $q$-independence
numerically
for all $n\leq 10$.

In addition to the CSW integrand~(\ref{allplusCSW}), one can also construct an equivalent ``BCFW-like'' integrand for the all-plus amplitude. In fact, it is easy to guess this alternative form of the integrand from the  all-$n$ expression
 for the tree amplitude
$\bigl\< \phi_1 \bar\phi_2\, g_3^+\ldots g_n^+\bigl\>$ of \cite{Ferrario:2006np,Craig:2011ws,Boels:2011zz} (see also~\cite{Forde:2005ue,Rodrigo:2005eu}).
It takes the form
\begin{equation}\label{WWallnceks}
    \bigl\< \phi_1 \bar\phi_2\, g_3^+\ldots g_n^+\bigl\>_\text{tree} ~=~
    -\frac{\mu^2}{\<34\>\<45\>\cdots \<n\!-\!1,n\>(P_{n1}^2+\mu^2)}\times
    \bigl[3\big|\prod_{j=3}^{n-2}\biggl[1+\frac{P_{J}|j\!+\!1\>[j\!+\!1|}{P_{J}^2+\mu^2}\biggr]\big|n\big]
    \,.
\end{equation}
This form of the amplitude was obtained using BCFW recursion relations.

 This suggests proceeding
as in the CSW case
 by $\Tr'$-ing
 the product in the BCFW-form \reef{WWallnceks}. This gives the following proposal for an alternative form of the all-plus integrand:
\begin{equation} \label{allplus2}
\boxed{
\begin{split}
  I_{\rm BCFW}^{++\cdots+}~=~\frac{2N_p}{\<12\>\cdots\<n1\>}\times
  \Tr'\prod_{j=1}^{n}\biggl[1+\frac{\ell_{j}|j\!+\!1\>[j\!+\!1|}{\ell_{j}^2+\mu^2}\biggr]\,.
\end{split}
}
\end{equation}
Indeed we have explicitly verified that the integrand~(\ref{allplus2}) correctly factorizes into the tree amplitude~(\ref{WWallnceks}) on the ``single cut'' of any  loop propagator $1/(\ell_i^2+\mu^2)$. Furthermore, we have numerically verified that
\begin{equation}
    I_{\rm BCFW}^{++\cdots+}(1,\ldots,n)~=~I_{\rm CSW}^{++\cdots+}(1,\ldots,n)
\end{equation}
for $n=3,4,\dots,10$. These two integrands are
thus expected to be literally identical, i.e.~not even differ
by terms that integrate to zero.\footnote{For example, at the four-point level, parity-odd terms with a numerator $\epsilon(p_1,p_2,p_3,\ell)$ integrate to zero because no four independent vectors are available to saturate the $\epsilon$-tensor.} We will therefore drop the label `CSW' or `BCFW' on the integrands $I^{++\cdots+}$ in the following.

Next we verify explicitly for
$n\leq 5$ that the all-plus integrand presented here is equivalent to the known expressions for the all-plus amplitude. Then we will move on to derive the one-minus integrand.

\vspace{2mm}
\noindent {\bf Explicit match to known expressions}\\
We have matched the all-plus integrand~(\ref{allplusCSW}),~(\ref{allplus2}) explicitly to   expressions in the literature for
$n\leq 5$. For the interested reader, the details   are given in appendix~\ref{appmatch}; here, we will briefly summarize the results.

To match to known expressions, it is convenient to start with the integrand in the BCFW representation~(\ref{allplus2}) and use the identity\footnote{
The subscript on $\Tr_\pm$ indicates that the trace is taken with a chiral projection $\frac{1}{2}(1\pm \gamma_5)$.}
\begin{equation}\label{matchID}
 \Tr'\prod_{j=1}^{n}\biggl[1+\frac{\ell_{j}|j\!+\!1\>[j\!+\!1|}{\ell_{j}^2+\mu^2}\biggr]
~=~
 \frac{ \Tr_-\big[(\ell_{1} \ell_{2} + \mu^2)\cdots(\ell_{n} \ell_{1} + \mu^2)\big]
 - \Tr_-\big[ d_1d_2 \cdots d_n \big]}
 {d_1d_2 \cdots d_n}\,,
\end{equation}
  with $d_i = \ell_i^2 + \mu^2$. For $n=3$, the two traces in~(\ref{matchID}) cancel, and
directly give
\begin{equation}\label{Ippp}
    I^{+++}(1,2,3) = 0\,,
\end{equation}
\emph{even before integration!} The vanishing of the all-plus 1-loop 3-point amplitude is of course well-known and thus anticipated.

Next we turn to the 4-point all-plus integrand.
We find
\bea\label{Ipppp}
  I^{++++}(1,2,3,4) ~\simeq~ 2N_p\,\frac{[12][34]}{\<12\>\<34\>} \frac{\mu^4}{d_1d_2d_3d_4}\,.
\eea
Here, `$\simeq$' signifies that we dropped parity-odd terms in the integrand which integrate to zero.
This result, the box integral for $I^{++++}$, is well-known in the literature~\cite{Bern:1995db}.

Finally, let us treat the $n=5$ case. This time we cannot discard the parity-odd contributions. Combining parity-even and non-vanishing parity-odd terms we  arrive at the following integrand
\bea
\label{Ippppp}
  I^{+++++} ~\simeq~
  \frac{2N_p}{\<12\>\<23\>\<34\>\<45\>\<51\>}
  \bigg(- \frac{1}{2}
  \bigg[
    \frac{\mu^4\,s_{12}s_{23}}{d_1d_2d_3d_5} + \text{cyclic}
  \bigg] + \frac{4i\mu^6\, \eps(1234) }{d_1d_2d_3d_4d_5}\bigg) \, .
\eea
The right-hand side is equivalent to the 5-point BCFW and CSW expressions
\reef{allplusCSW} and \reef{allplus2} for $n=5$ after dropping  several parity-odd terms that integrate to zero, as explained in more detail in appendix~\ref{appmatch}.
 The result \reef{Ippppp} is a sum of five box integrals and a pentagon integral; this form is known in the literature   \cite{Bern:1995db}.  Thus we have shown that for $n=3,4,5$ our integrand reproduces the known amplitudes.

\subsection{The one-minus integrand}

\begin{figure}
\begin{center}
\includegraphics[height=3.5cm]{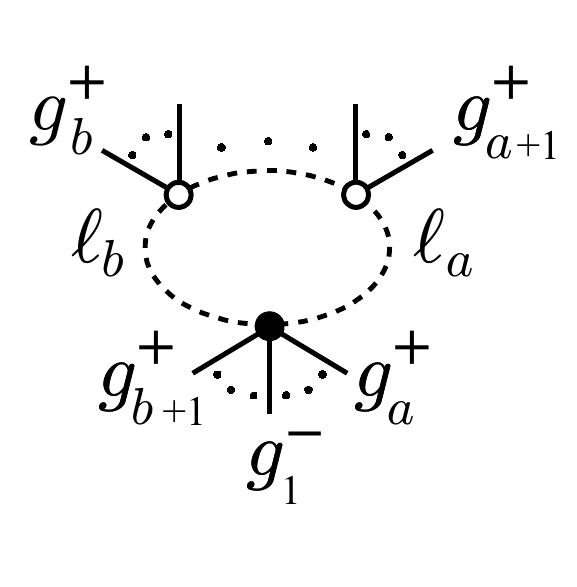}
\hskip.9cm
\includegraphics[height=3.5cm]{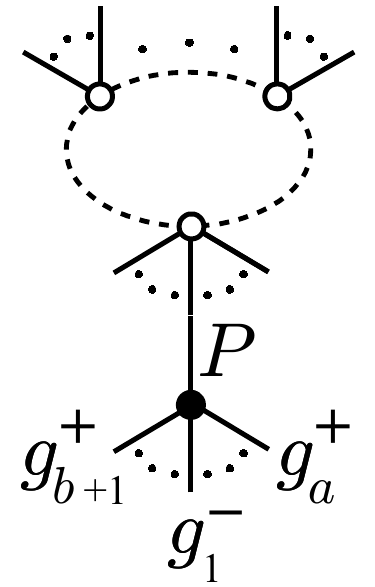}
\hskip1.3cm
\includegraphics[height=3.5cm]{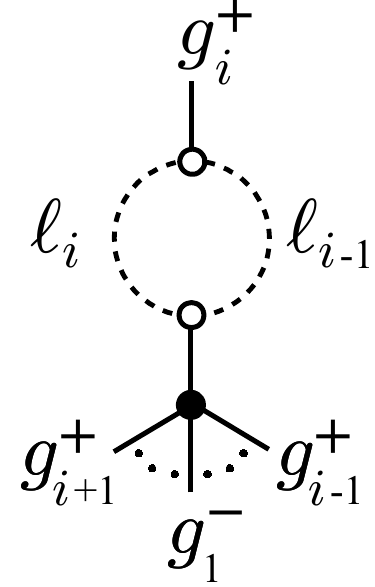}
\hskip1.6cm
\includegraphics[height=3.5cm]{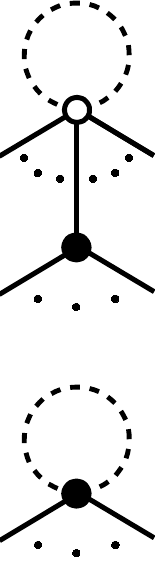}
\end{center}
\vskip-.05cm
\hspace{3.05cm}(i)\hskip3.5cm(ii)\hskip3.103cm(iii)\hskip2.7cm(iv)
\caption{The diagrams of the one-minus rational integrand: the
finite ring (i) and subtree (ii) contributions, and the divergent external bubble (iii) and tadpole diagrams (iv). The divergent diagrams are dropped and replaced by the correction term
$I^{-+\cdots+}_{\rm sprs}$.
}
\label{figdiags}
\end{figure}

Let us now consider the integrand of the ``one-minus'' amplitude, $I_n^{-+\cdots+}$, in QCD.
This  amplitude vanishes in supersymmetric gauge theories so,  like the all-plus amplitude,  it only receives contributions from the scalar  loop in the decomposition~(\ref{decomp}).

Naively, all the diagrams we need to consider for the one-minus integrand are given in figure \ref{figdiags}. However, some of these diagrams are divergent, namely the external bubble diagrams
 in figure
\ref{figdiags}(iii) and
 some of the tadpoles diagrams in figure
\ref{figdiags}(iv).
The tadpole contributions are harmless as argued in~\cite{Brandhuber:2007vm}, and we will simply drop them. Our analysis below verifies that no tadpole-like correction terms need to be added to the integrand to ensure $q$-independence.
The external bubble diagrams \ref{figdiags}(iii) involve bubbles on external lines, and they are divergent because they involve an on-shell internal propagator.
Thus we have to be more careful, and we now discuss the approach.

\subsubsection{External bubble contributions}
Unlike the all-plus integrand, the computation of the one-minus integrand faces a major obstacle: the one-minus integrand receives contributions from
diagrams with an external massless bubble.
These external bubble diagrams are divergent and must be ``amputated''. In conventional gauges, say Feynman gauge, this amputation is straight-forward because external bubbles correspond to massless bubble integrals that integrate to zero in dimensional regularization. Amputation thus simply amounts to dropping all diagrams with bubbles on the external lines. In the   CSW diagrammatic rules, however, the external bubble diagrams depicted in
Figure~\ref{figdiags}(iii)
contain spurious poles in the loop momentum of the form $1/\<i|\ell_i|q]$, and therefore do not necessarily
integrate to zero. As a consequence, naively dropping all divergent tadpole and external-bubble contributions gives a wrong integrand that contains spurious poles.

To deal with this problem, we follow a two-step strategy:
\begin{enumerate}
  \item[{\bf Step 1:}~] We first write down the naive integrand $I^{-+\cdots+}_{\rm naive}$ that is simply the sum of all finite, non-divergent diagrams contributing to the  CSW expansion of the integrand. These diagrams are illustrated in Figure~\ref{figdiags}(i)  and (ii).  In the CSW expansion, all diagrams are finite if they contain at least two propagators of loop momenta $\ell_i$, $\ell_j$ that are non-adjacent, $j\neq i\pm1$. Therefore, $I^{-+\cdots+}_{\rm naive}$ correctly reproduces all ($D$-dimensional) bubble cuts of two such non-adjacent loop momenta. In particular, all triangle, box and pentagon cuts are also correctly reproduced from $I^{-+\cdots+}_{\rm naive}$. However, it still contains spurious poles;   these are present in cuts of two adjacent loop momenta, $\ell_{i-1}$ and $\ell_i$.
  \item[{\bf Step 2:}~] We determine a correction term $I^{-+\cdots+}_{\rm sprs}$ that satisfies two crucial properties:
  \begin{itemize}
    \item $I^{-+\cdots+}_{\rm sprs}$ removes the spurious $q$-dependence from $I^{-+\cdots+}_{\rm naive}$,  so that $I^{-+\cdots+}_{\rm naive}\!\!+I^{-+\cdots+}_{\rm sprs}$ is independent of $q$ and thus free of spurious poles.
    \item $I^{-+\cdots+}_{\rm sprs}$ vanishes on any cut of two non-adjacent loop propagators; therefore,  $I^{-+\cdots+}_{\rm sprs}$ can be written as a sum over terms that each contain two adjacent loop propagators, $\sim 1/{(\ell_{i-1}^2+\mu^2)(\ell_i^2+\mu^2)}$.
  \end{itemize}
  $I^{-+\cdots+}_{\rm sprs}$ should be interpreted as the finite parts hidden in the divergent external-bubble diagrams of Figure~\ref{figdiags}(iii) that are needed to render the integrand $q$-independent.
\end{enumerate}
Below, we will determine a $I^{-+\cdots+}_{\rm sprs}$ with these properties. We then claim that
\begin{equation}
    I_n^{-+\cdots+}~\equiv~ I^{-+\cdots+}_{\rm naive}+I^{-+\cdots+}_{\rm sprs}
\end{equation}
is the correct integrand of the one-minus amplitude. Indeed, $I_n^{-+\cdots+}$ only contains physical poles and factorizes correctly on all $D$-dimensional bubble cuts of two loop momenta $\ell_i$, $\ell_j$ that are non-adjacent, $j\neq i\pm1$. In the absence of spurious poles, the only remaining ambiguity are terms proportional to adjacent-line
bubble and tadpole integrals; but such $D$-dimensional integrals have no  scale and vanish in dimensional regularization! It follows that $I_n^{-+\cdots+}$ determined by the two-step strategy gives the correct one-minus amplitude.

 \subsubsection{Explicit all-$n$ integrand}
We now carry out the two-step strategy explicitly to determine $I_n^{-+\cdots+}$  for any number of external legs $n$.

{\bf Step 1} is straight-forward; there are two types of finite diagrams contributing to $I_n^{-+\cdots+}$. The {\em ring diagrams}, which are schematically displayed in Figure~\ref{figdiags}(i),  consist of a ring of vertices, all of which are UHV except for one MHV vertex containing the negative-helicity line 1. The entire contribution from ring diagrams can be combined into the following compact expression:
\begin{equation}
\begin{split}
  I^{-+\cdots+}_{\rm ring}(1,\ldots,n)~
  &=~
  2N_p\sum_{b>a}
  \frac{-\mu^2\<1\ell_a\>^2\<1\ell_b\>^2\<a,a\!+\!1\>\<b,b\!+\!1\>}
  {\<12\>\cdots\<n1\>\<\ell_a\ell_b\>\<a\ell_a\>\<\ell_a,a\!+\!1\>\<b,\ell_b\>\<\ell_b,b\!+\!1\>(\ell_a^2+\mu^2)(\ell_b^2+\mu^2)}\\
  &\hskip2.7cm\times
  \<\ell_{a}|\prod_{j=a+1}^{b-1}\biggl[1-\frac{\mu^2|\ell_j\>\<j,j\!+\!1\>\<\ell_j|}{(\ell_j^2+\mu^2)\<\ell_j,j\>\<j\!+\!1,\ell_j\>}\biggr]|\ell_b\>\,.
\end{split}
\end{equation}
Expanding the product over $j$ reproduces each individual ring diagram in the   CSW expansion.

The second contribution comes from {\em subtree diagrams}, consisting  of a ring of UHV vertices
that is connected via a propagator $1/P^2$ to an MHV vertex that contains line 1. This contribution is illustrated in Figure~\ref{figdiags}(ii). The computation of the ``ring'' part of these diagrams coincides with our analysis for the all-plus integrand  $I^{++\cdots+}$ in section~\ref{secallplus}. We find,
\begin{equation}
\begin{split}
  I^{-+\cdots+}_{\rm subtree}(1,\ldots,n)
  = &\sum_{2\leq b-a\leq n-2}
  \frac{\<1P\>^4}{\<P,b\!+\!1\>\<b\!+\!1,b\!+\!2\>\cdots\<a\!-\!1,a\>\<aP\>}
 \times\frac{1}{P^2}\times  I_{\rm CSW}^{++\cdots+}(a\!+\!1,\ldots,b,P)\,.
\end{split}
\end{equation}
where
\begin{equation}
    P\equiv p_{a+1}+p_{a+2}+\dots+p_{b}\,.
\end{equation}
 The range of $a$ and $b$ in the sum is chosen   such that  $a$ and $b$ are non-adjacent.
As $P$ is an off-shell momentum, the CSW prescription is understood for all occurrences of $|P\>$ in the CSW all-plus integrand $I_{\rm CSW}^{++\cdots+}$, defined in~(\ref{allplusCSW}).
The naive integrand is the sum of the ring and subtree contributions,
\begin{equation}
    I^{-+\cdots+}_{\rm naive}~=~I^{-+\cdots+}_{\rm ring}+I^{-+\cdots+}_{\rm subtree}\,.
\end{equation}
As it stands, the integrand $ I^{-+\cdots+}_{\rm naive}$ factorizes correctly on $D$-dimensional pentagon, box, triangle, and non-adjacent bubble cuts. However, it contains uncanceled spurious singularities of the form $1/\<i\ell_i\>^2$ and $1/(\<i-1,\ell_{i-1}\>\<i\,\ell_i\>)$ with $i=2,\ldots,n$\,, where the $|\ell_i\>$ depend on the reference $q$ through the CSW prescription~(\ref{CSW}). This is not surprising, considering that we have dropped the (divergent) external bubble contributions displayed in Figure~\ref{figdiags}(iii)
that contain such spurious singularities.

We now proceed with {\bf step 2} of the above strategy, and try to determine a correction term $I^{-+\cdots+}_{\rm sprs}$ that cancels the spurious $q$-dependence in $I^{-+\cdots+}_{\rm naive}$ without spoiling its crucial factorization properties. We make the ansatz
\begin{equation}
    I^{-+\cdots+}_{\rm sprs}~=~2N_p\sum_{i=2}^n\frac{\mu^2}{\<12\>\cdots\<n1\>(\ell_{i-1}^2+\mu^2)(\ell_i^2+\mu^2)}
    \Biggl(\frac{D_i}{\<i\ell_i\>^2}+\frac{S_i}{\<i\ell_i\>}\Biggr)\,,
\end{equation}
where the residues of the double and single
spurious poles in
 $\<i\ell_i\>\!=\!\<i|\ell_i|q]$
are controlled by the kinematic coefficients $D_i$ and $S_i$. These coefficients are highly constrained by little-group properties and are not allowed to contain any $\ell$-dependent denominator factors. A numeric analysis gives the following solution:
\begin{equation}
    D_i~=\,-\<1i\>^2\<1\ell_{i-1}\>\<1\ell_i\>\,,~\qquad
    S_i~=~\<1i\>^2\biggl[\frac{\<1,i-1\>\<1\ell_i\>}{\<i-1,i\>}
        -\frac{\<1,i+1\>\<1\ell_{i-1}\>}{\<i,i+1\>}\biggr]\,.
\end{equation}
While not obvious, $I^{-+\cdots+}_{\rm sprs}$ indeed cancels all spurious poles in the naive integrand, rendering it $q$-independent.\footnote{We could of course shift $I^{-+\cdots+}_{\rm sprs}$ by any $q$-independent function that does not spoil factorization properties, \eg we could shift $D_i\to D_i+f(p_i,\ell_i)\<i\ell_i\>^2$, where $f(p_i,\ell_i)$ is a function of of $p_i$ and $\ell_i$, with only polynomial dependence on the $\ell_i$.  However, such a shift term is proportional to a scaleless integral and thus does not affect the amplitude. It integrates to zero.}
In summary, the $n$-point one-minus integrand is given by
\begin{equation}\label{oneminus}
\boxed{
\begin{split}
  &I_n^{-+\cdots+}
  ~=~I^{-+\cdots+}_{\rm ring}+I^{-+\cdots+}_{\rm subtree}+ I^{-+\cdots+}_{\rm sprs}\phantom{\biggl(}\\
    &=~\!\!\!\sum_{b>a}\frac{-2N_p\,\mu^2\<1\ell_a\>^2\<1\ell_b\>^2\<a,a\!+\!1\>\<b,b\!+\!1\>}
  {\<12\>\cdots\<n1\>\<\ell_a\ell_b\>\<a\ell_a\>\<\ell_a,a\!+\!1\>\<b,\ell_b\>\<\ell_b,b\!+\!1\>(\ell_a^2+\mu^2)(\ell_b^2+\mu^2)}\\
  &\hskip2.7cm\times
  \<\ell_{a}|\prod_{j=a+1}^{b-1}\biggl[1-\frac{\mu^2|\ell_j\>\<j,j\!+\!1\>\<\ell_j|}{(\ell_j^2+\mu^2)\<\ell_j,j\>\<j\!+\!1,\ell_j\>}\biggr]|\ell_b\>\\
  &~~~~+\sum_{2\leq b-a\leq n-2}
  \frac{\<1P\>^4}{\<P,b\!+\!1\>\<b\!+\!1,b\!+\!2\>\cdots\<a\!-\!1,a\>\<aP\>}
 \times\frac{1}{P^2}\times  I_{\rm CSW}^{++\cdots+}(a\!+\!1,\ldots,b,P)\\
 &~~~~+
    \sum_{i=2}^n
    \frac{-2N_p\,\mu^2\<1i\>^2}{\<12\>\cdots\<n1\>(\ell_i^2\!+\!\mu^2)(\ell_{i\dash1}^2\!+\!\mu^2)\<i\,\ell_i\>}
    \times
    \Biggl[
    \frac{\<1\ell_{i\dash1}\>\<1\ell_i\>}{\<i \ell_i\>}
    \!-\!\frac{\<1,i\!-\!1\>\<1\ell_i\>}{\<i\!-\!1,i\>}
    \!+\!\frac{\<1,i\!+\!1\>\<1\ell_{i\dash1}\>}{\<i,i\!+\!1\>}
    \Biggr]\,.
\end{split}
}
\end{equation}

We have performed various checks on the correctness of the integrand~(\ref{oneminus}). Specifically, we  have numerically verified $q$ independence of the integrand for $n=4,5,6,7,8,9,10$. For $n=4$, we have gone further and explicitly re-expressed the integrand $I_4^{-+++}$ in a manifestly $q$-independent form. We have then performed integral reduction on this form and matched it to the result of Bern and Morgan~\cite{Bern:1995db},
\begin{equation}
\begin{split}\label{bm}
 &A_4^{\rm QCD}(1^-,\, 2^+, 3^+, 4^+) \\
 &=
  {2iN_p \over (4\pi)^{2-\eps}}
  { [24]^2 \over [12] \<23\>\< 34\> [41] }
 \frac { s t }{ u } \biggl[
      \frac{ t(u\,\dash\, s)}{  s u } J_3(s)
      + \frac{ s(u\,\dash\, t)}{  t u } J_3(t)
      - { t\,\dash\, u \over s^2 } J_2(s)
      - { s\,\dash\, u \over t^2 }J_2(t)
      + { s t \over 2 u } J_4
      + K_4
  \biggr] \, .
 \end{split}
\end{equation}
Here, $K_4$ is a box integral in $D=8-2\epsilon$ dimensions, while $J_2$, $J_3$ and $J_4$ are bubble, triangle and box integrals in $D=6-2\epsilon$ dimensions (see~\cite{Bern:1995db} for a precise definition).    To match $I_4^{-+++}$ to the integrand in~(\ref{bm}), we  dropped terms that integrate to zero.

\setcounter{equation}{0}
\section{CSW expansion for Coulomb-branch amplitudes in $\cn=4$ SYM}\label{secCoulomb}
In this section, we derive the massive   CSW expansion for Coulomb-branch amplitudes in $\cn=4$ SYM that was proposed in~\cite{Kiermaier:2011cr}. We first briefly review $\cn=4$ SYM theory on the Coulomb branch and the proposed CSW expansion. We then examine the supersymmetric properties of the massive CSW rules. Finally, we present a proof of the expansion.

\subsection{Review:  $\cn=4$ SYM on the Coulomb branch and its   massive CSW expansion}\label{secCoulombReview}
We consider $\cn\!=\!4$ SYM with gauge group $U(M\!+\!N)$. The simplest way to move onto the Coulomb-branch is to give vevs to a subset of the scalars,
\begin{equation}
    \big\<(\phi_{12})_A{}^{B}\big\>\, = \,\big\<(\phi_{34})_A{}^{B}\big\>  \,=\, m \, \delta_{A}{}^{B}\qquad \qquad\text{ for }~~~ 1\leq A,B\leq M\,.
\end{equation}
Here and in the following we suppress all coupling dependence,
effectively setting $g=1$.
These vevs break the gauge group spontaneously to $U(M)\!\times\! U(N)$, and the $R$-symmetry group as $SU(4)\to Sp(4)$.
 They also split
the states into a massless and a massive sector.
The massless sector contains the gluons $g^\pm$, fermions $\chi^a$, and scalars $\phi^{ab}$, where $a,b$ are R-symmetry indices. The massive sector contains fields of mass $m$ that are bifundamental with respect to $U(M)\!\times\! U(N)$, consisting of $W$ bosons, scalars $w$, and fermions
 $\Psi$.
The conjugate particles in the bifundamental of $U(N)\!\times\! U(M)$ have mass
 parameter
$-m$.
Table~\ref{tabstates} summarizes the massless and massive states, their polarizations and wave functions, and how they correspond to each other.
\begin{table}[t!]
\begin{center}
\begin{tabular}{r ccc }
&massless fields  & massive fields &
wave functions
\\
\noalign{\smallskip}
\cline{2-4}\\[-2.5ex]
\cline{2-4}
\noalign{\medskip}
gluons / $W^\pm$-boson: & $g^+$, $g^-$ &  $W^+$, $W^-$
&$\epsilon_-\!=\!\frac{\sqrt{2}|i^\perp\> [q|}{[ i^\perp q]},~
     \epsilon_+\!=\!\frac{\sqrt{2}|q\> [i^\perp|}{\< i^\perp q\>}$
\\
\myhline
scalar / $W^L$-boson: & $\tfrac{1}{\sqrt{2}}(\phi^{12}\!+\!\phi^{34})$
&  ~$W^L\!\sim\! \tfrac{1}{\sqrt{2}}(w^{12}\!+\!w^{34})$~
&
$\slashed{\epsilon}_L =
    \frac{1}{m_i} \Big( \slashed{p}_i^\perp  + \frac{m_i^2}{2q\cdot
p_i} \slashed{q} \Big)$
\\
\myhline
\multirow{2}{*}{scalars:} & ~~$\phi^{13}$\!, $\phi^{14}$\!, $\phi^{23}$\!,
$\phi^{24}$\!, & $w^{13}$, $w^{14}$, $w^{23}$, $w^{24}$,
&\multirow{2}{*}{}\\[.3ex]
&$\tfrac{1}{\sqrt{2}}(\phi^{12}\!-\!\phi^{34})$&$\tfrac{1}{\sqrt{2}}(w^{12}\!-\!w^{34})$\\
\myhline
fermions:& $\chi^a$, $\chi^{abc}$  &
$\Psi^a$
&
$v_+\!=\!\begin{pmatrix} \!|p^\perp]\\[1mm]\! \frac{im|q\>}{\<qp^\perp\>} \end{pmatrix},~  v_-\!=\!
    \begin{pmatrix} \!\frac{im|q]}{[p^\perp q]}\\[1mm]\!|p^\perp\> \end{pmatrix}$
\end{tabular}
\caption{\small Massless and massive particles  on the Coulomb branch for
the R-symmetry breaking~$SU(4)\!\to\!Sp(4)$.
The massive fermions $\Psi^a$ are 4-component Dirac fermions.
}
\end{center}
\label{tabstates}
\end{table}

The Coulomb branch of $\cn=4$ SYM can be interpreted as arising from dimensional reduction of massless $\cn=(1,1)$ SYM in 6 dimensions. In this interpretation, the mass parameters $m_i$ of particles are
 related to  momenta in the extra dimensions, $m_i=p_5+ip_6$.
The external particles of any non-vanishing Coulomb-branch amplitude must satisfy
\begin{equation}\label{summi}
    \sum_im_i~=~0\,.
\end{equation}
For simplicity, we will take the $m_i$ to be real (but either +ve or -ve) in the following  and refer to them  as ``masses''.

\vspace{2mm}
\noindent {\bf The massive spinor-helicity formalism}\\
A convenient way to express amplitudes on the Coulomb-branch of $\cn=4$ SYM is the massive spinor-helicity formalism~\cite{Kleiss:1985yh,Dittmaier:1998nn}.\footnote{
We use the conventions in~\cite{Cohen:2010mi,Bianchi:2008pu}.}
 One  decomposes  a massive on-shell momentum $p_i$ of mass $m_i$ in terms of a pair of null vectors, a  reference null $q$ and the null projection $p_i^\perp$, viz.
\be \lab{massivep}
p_i = p_i^\perp - \frac{m_i^2}{2q\cdot p_i} \,q\,,\qquad\quad
p_i^2 = - m_i^2\,.
\ee
Since $p_i^\perp$ and $q$ are null vectors, there are associated spinors
$|i^\perp\>, |i^\perp] ,~|q\>,|q]$, such that
\bea
(p_i^\perp)^{\dot\alpha \alpha}~=~|i^\perp\>^{\dot\alpha}[i^\perp|^{\alpha}\,,\qquad
q^{\dot\alpha \alpha}~=~|q\>^{\dot\alpha}[q|^{\alpha}\,.
\eea
For massive vector bosons, the spinors $|q\>$ and $|q]$ allow us to define a convenient basis of polarization vectors:
\bea\label{pol}
    \epsilon_-=\frac{\sqrt{2}|i^\perp\> [q|}{[ i^\perp q]}~~\,,~\qquad
    \epsilon_+=\frac{\sqrt{2}|q\> [i^\perp|}{\< i^\perp q\>}
    ~~\,,~\qquad
  \slashed{\epsilon}_0 &=&
   \frac{1}{m_i} \Big( \slashed{p}_i^\perp  - \frac{m_i^2}{\<q|p_i|q]} \slashed{q} \Big)\,.
\eea
In the following we will denote this basis of polarization vectors as ``$q$-helicity basis''. For example, vector bosons with polarizations $\epsilon_\pm$ and $\epsilon_0$ have $q$-helicity $h=\pm 1$ and $h=0$, respectively.
It is convenient to use the spinor $|q]$ also as the reference spinor in the CSW expansion.

\vspace{2mm}
\noindent {\bf MHV-classification}\\
The familiar N$^k$MHV classification of massless $\mathcal{N}=4$ SYM amplitudes has to be augmented when applied to
Coulomb branch amplitudes. Each of the two $SU(2)$-sectors of the unbroken $Sp(4)$ R-symmetry has an N$^k$MHV classification with non-vanishing amplitudes for $k=-1$ (ultra-helicity violating, UHV), $k=0$ (MHV), $k=1$ (NMHV) etc.\footnote{For the case of massless amplitudes in  $\mathcal{N}=4$ SYM, the amplitudes with $k=-2,-1$ vanish; they correspond to the sectors  of all-plus or one-minus amplitudes. In
 the
massive spinor helicity formalism where the same reference vector $q$ is used for all states, the all-plus amplitudes $k=-2$
 still
vanish, but the UHV amplitudes with $k=-1$ are non-vanishing.}   Thus we classify the massive Coulomb-branch amplitudes as UHV$\times$UHV, UHV$\times$MHV, MHV$\times$MHV etc. When no confusion
 is possible,
we will
refer to UHV$\times$UHV and MHV$\times$MHV as the UHV and MHV sectors, respectively.

\vspace{2mm}
\noindent {\bf Soft-limit construction of massive amplitudes from massless amplitudes}\\
In~\cite{Craig:2011ws,Kiermaier:2011cr}, it was proposed that massive Coulomb-branch on-shell amplitudes can be expressed in terms of massless amplitudes at the origin of moduli space. Non-trivial evidence for this proposal was presented in~\cite{Craig:2011ws} at leading order, and in~\cite{Kiermaier:2011cr} to all orders. We now review the details of this proposal, for the special case of Coulomb-branch tree-level scattering of two adjacent massive $W$-bosons $W_1$, $W_2$ with an arbitrary number of additional massless particles. Such an amplitude can be expressed in terms of massless $\cn=4$ amplitudes as
\begin{equation}\label{proposal}
    \bigl\<\,W_1\barW_2\,\ldots\,\bigr\>~~=~~\lim_{\varepsilon\to 0}\,\sum_{s=0}^\infty\,\, \bigl\<\,g_1\,\underbrace{\phi^{\rm vev}_{\varepsilon q}\ldotss\phi^{\rm vev}_{\varepsilon q}}_{s
    \text{ times}}\,g_2\,\ldots\,\bigr\>_{\rm sym}\,,
\end{equation}
where the $\ldots$ represent arbitrary further massless particles in the amplitude.
Some elaborations on the proposal~(\ref{proposal}) are in order:
\begin{itemize}
  \item The polarizations of the $W$-bosons on the left-hand side are chosen in the $q$-helicity basis~(\ref{pol}).
  The massless gluons $g_1$, $g_2$ have the corresponding massless helicity.
  \item The massless gluons $g_1$, $g_2$ on the right-hand side have momenta $p_i^\perp$ that are related to the massive momenta of the $W$-bosons via~(\ref{massivep}).
  \item The reference vector $q$ is subject to the constraint
\begin{equation}\label{spq}
    \sum_{i=1}^n\frac{m_i^2}{2\,q\!\cdot\! p_i}~=~0\,,
\end{equation}
 which ensures momentum conservation on the right-hand side, $\sum_i p_i^\perp=0$. For the two-mass case at hand,~(\ref{spq}) is equivalent to the simple orthogonality condition $q\cdot (p_1+p_2)=0$\,.
  \item The scalar $\phi^{\rm vev}_{\varepsilon q_i}$ is a massless soft scalar
  of momentum $\varepsilon q_i$ whose R-symmetry structure is oriented in the Coulomb-branch vev direction, $\phi^{\rm vev}=\<\phi_{ab}\>\phi^{ab}$. In our case, we thus have
  \begin{equation}
    \phi^{\rm vev}~=~m\bigl(\phi^{12}+\phi^{34}\bigr)\,.
  \end{equation}
  \item
 The subscript `sym' denotes a symmetrization of the vev scalars in their
  momenta
 $q_i$ before taking the collinear limit $q_i\to q$. This sum over permutations ensures that they are ``unordered'' particles in the massless partial amplitudes, which befits a vev scalar that must live in the Cartan subalgebra and thus commute with itself.
 The symmetrization ensures that the the right-hand side of~(\ref{proposal}) is finite in the collinear limit $q_i\to q$.
\end{itemize}
It was shown in~\cite{Kiermaier:2011cr}, that the multi-soft limit in~(\ref{proposal}) is well-defined,
\ie it is free of collinear and soft divergences. It was also shown that the proposal~(\ref{proposal}), and its generalization to amplitudes and loop integrands with arbitrarily many massive particles, implies a   massive CSW vertex expansion, which we now review.

\vspace{2mm}
\noindent {\bf  Massive CSW rules}\\
In~\cite{Kiermaier:2011cr}, it was shown that the soft-limit construction detailed above is equivalent to a   massive CSW expansion for Coulomb-branch
amplitudes in the $q$-helicity basis. We now review the diagrammatic rules of this expansion.

The propagators in the   massive CSW expansion are conventional massive scalar propagators:
\begin{equation}\label{propcsw}
        \parbox[c]{1.8cm}{\includegraphics[width=1.6cm]{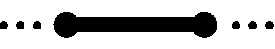}}~=~\frac{1}{P_I^2+m_I^2}\,.
\end{equation}
 Just like momentum is conserved at each vertex, the mass parameters $m_i$ also sum to zero at each vertex;
therefore, the internal mass $m_I$ is given by the sum of masses of the other lines at the left or right vertex. Of course,~(\ref{propcsw}) includes massless propagators as a special case when $m_I\!=\!0$.

\noindent There are three types of vertices in the expansion:\\[-7mm]
\begin{itemize}
\item
The first vertex is the conventional MHV vertex, with perp'ed spinors:
\begin{equation}\label{vert1}
    \parbox[c]{2.2cm}{\includegraphics[width=1.5cm]{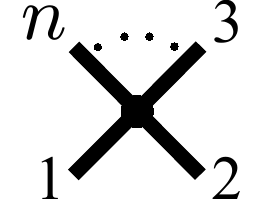}}~=~\frac{\delta^{(8)}\big(|i^\perp\>\eta_{ia}\big)}{\<1^\perp2^\perp\>\cdots\<n^\perp1^\perp\>}\,.
\end{equation}
\item
The second vertex is an ultra-helicity-violating (UHV) vertex:
\begin{equation}\label{vert2}
        \parbox[c]{1.7cm}{\includegraphics[width=1.5cm]{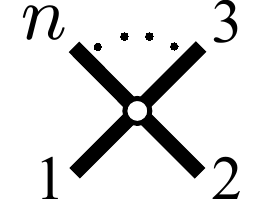}}~=~K_n^2\times\frac{\delta^{(4)}\big(\<qi^\perp \>\eta_{ia}\big)}{\<1^\perp2^\perp\>\cdots\<n^\perp1^\perp\>}\,.
\end{equation}
The kinematic prefactor $K_n$ is given by
\begin{equation}
 \label{Kn}
    K_n~=~\sum_{i}\frac{m_i\<X i^\perp\>}{\<X q\>\<i^\perp q\>}\,,
\end{equation}
for arbitrary reference spinor $|X\> \ne |q\>$.
In fact, using $\sum_i m_i=0$ it is easy to see that $K_n$ is independent of the choice of $|X\>$~\cite{Kiermaier:2011cr}.
The vertex \reef{vert2} is $O(m^2)$ and thus not present for massless amplitudes.

The vertex~(\ref{vert2})  generalizes the UHV vertices~(\ref{UHVsc}) that we encountered in the scalar-vector theory; in fact, a short computation shows that, with $m\equiv m_1=-m_2$ and $m_i=0$ for $i\geq3$, we can reproduce the vertex~(\ref{UHVsc}) by projecting out a pair of conjugate scalars on lines $1$ and $2$:
\begin{equation}
    \frac{\partial^2}{\partial \eta_{11}\partial \eta_{12}}\,\frac{\partial^2}{\partial \eta_{23}\partial \eta_{24}}\,\,\frac{K_n^2\,\delta^{(4)}\big(\<qi^\perp \>\eta_{ia}\big)}{\<1^\perp2^\perp\>\cdots\<n^\perp1^\perp\>}~=~\frac{m^2\<12\>}{\<23\>\cdots\<n1\>}\,.
\end{equation}
In particular, all dependence on the holomorphic reference spinor $|q\>$ cancels in this case.

\item Finally, there is a third vertex, which breaks the R-symmetry $SU(4)\to Sp(4)$. This ``MHV$\times$UHV vertex'' has the structure of the MHV vertex~(\ref{vert1}) with respect to one of the two $SU(2)$ factors in $Sp(4)$, and the structure of the UHV vertex~(\ref{vert2}) with respect to the other $SU(2)$. Explicitly,
\begin{equation}\label{vert3}
        \parbox[c]{1.7cm}{\includegraphics[width=1.5cm]{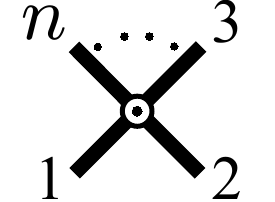}}~=~K_n\times\frac{\delta^{(4)}_{12}\big(|i^\perp\>\eta_{ia}\big)\,\delta^{(2)}_{34}\big(\<qi^\perp\>\eta_{ia}\big)
    +\delta^{(4)}_{34}\big(|i^\perp\>\eta_{ia}\big)\,\delta^{(2)}_{12}\big(\<qi^\perp\>\eta_{ia}\big)
    }{\<1^\perp2^\perp\>\cdots\<n^\perp1^\perp\>}\,,
\end{equation}
where $K_n$ is given by \reef{Kn}.
The subscripts on the Grassmann $\delta$-functions  indicate  which of the two $SU(2)$ factors of $Sp(4)$ the $\delta$-function `lives in'.
\end{itemize}

An example for an amplitude computed from these rules is the scattering of two $W$ bosons of mass $m$ with arbitrarily many massless gluons $g$,
\begin{equation}\label{WWalln}
    \bigl\< W_1^- \barW_2^+\, g_3^+\ldots g_n^+\bigl\>~=~-
    \frac{m^2\<q1^\perp\>^2}{\<q2^\perp\>^2\<2^\perp 3\>\<34\>\cdots \<n1^\perp\>}\times \big\<2^\perp\big|\prod_{j=3}^{n-1}\biggl[1-\frac{m^2|P_J\>\<j,j\!+\!1\>\<P_J|}{(P_J^2+m^2)\<P_J,j\>\<j\!+\!1,P_J\>}\biggr]
    \big|1^\perp\big\>\,,
\end{equation}
where we denoted $P_J\equiv P_{2..j}$.
This amplitude is related by supersymmetry to the massive-scalar amplitude~(\ref{WWalln1}). In fact, these two amplitude only differ by the spinor factor
$\<q1^\perp\>^2/\<q2^\perp\>^2$, which corrects the helicity weights. The amplitude~(\ref{WWalln}) has only one negative-helicity particle and is thus in the UHV sector. UHV amplitudes vanish in the massless limit due to supersymmetry. The UHV sector is the ``lowest'' non-vanishing sector on the Coulomb branch. Indeed, unlike in non-supersymmetric theories, the all-plus amplitude vanishes in $\cn=4$ SYM even on the Coulomb branch. This follows directly from SUSY Ward identities~\cite{Boels:2007pj}.

For $q$'s satisfying~(\ref{spq}), it was shown in~\cite{Kiermaier:2011cr} that this  massive CSW expansion is identical to the soft-limit construction~(\ref{proposal}).
In
section~\ref{secder}, we will present a recursion relation derivation of this expansion that is in fact valid for {\em any} choice of reference vector $q$.

\subsection{Manifest $\tQ$-supersymmetry of  massive CSW rules}\label{sectQ}
The  CSW rules in {\em massless} $\cn=4$ SYM \ manifestly preserve the $\tQ$ supercharges, diagram by diagram. In fact, MHV vertices contain an overall factor of
$\delta^{(8)}(|\tQ\>)$, where $|\tQ_{a}\>$ are the holomorphic supercharges in the massless case, $|\tQ_{a}\>=\sum_i|i\>\eta_{ia}$.
On the Coulomb branch, these supercharges are deformed because the super-algebra acquires a central charge:
\begin{equation}\label{tQsusy}
    |\tQ_{a}\>\,=\,\sum_i|\tQ_{ia}\>\,,\qquad\text{ with }~~~~
    |\tQ_{ia}\>\,=\,|i^\perp\>\eta_{ia}-\frac{m_i|q\>}{\<q i^\perp \>}\Omega_{ab}\frac{\partial}{\partial\eta_{ib}}\,, \quad
    \Omega_{ab}~=~
  \begin{pmatrix}
     i\sigma_2 & 0 \\
    0 &  i\sigma_2
  \end{pmatrix} \,.
\end{equation}
For example,
\begin{equation}
        |\tQ_{i1}\>\,=\,|i^\perp\>\eta_{i1}-\frac{m_i|q\>}{\<q i^\perp \>}\frac{\partial}{\partial\eta_{i2}}\,.
\end{equation}
Each vertex in the  massive CSW expansion,~(\ref{vert1}),~(\ref{vert2}), and~(\ref{vert3}),  is individually invariant only under half of the $\tQ$ SUSY charges, namely under the $q$-projection of $|\tQ_{a}\>$,
\begin{equation}
    \<q\tQ_{a}\>=\sum_i\<qi^\perp\>\eta_{ia}\,.
\end{equation}
They cannot be individually invariant under the entire $|\tQ_{a}\>$-symmetry, because the full generators~(\ref{tQsusy}) mix different $\eta$ degrees.
It is therefore convenient to combine vertices of different $\eta$-degree into a
\emph{supervertex},\footnote{We thank C. Peng for pointing out that
 the
supervertex factorizes and is useful in explicit calculations.}
\begin{equation}
\label{superV}
\begin{split}
    \parbox[c]{1.5cm}{\includegraphics[width=1.5cm]{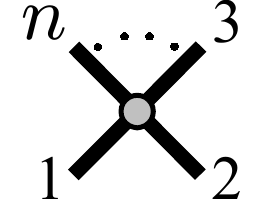}}~&\equiv~
    \parbox[c]{1.5cm}{\includegraphics[width=1.5cm]{cswnpt2softgen}}
    +    \parbox[c]{1.5cm}{\includegraphics[width=1.5cm]{cswnpt1softgen}}+
        \parbox[c]{1.5cm}{\includegraphics[width=1.5cm]{cswnptgen}}
        ~=~
    \frac{{\cal V}_n
    }{\<1^\perp2^\perp\>\cdots\<n^\perp1^\perp\>}\,,
\end{split}
\end{equation}
with
\begin{equation}\label{Vn}
    {\cal V}_n~=~
    \Bigl[\delta^{(4)}_{12}\big(|i^\perp\>\eta_{ia}\big)+K_n\delta^{(2)}_{12}\big(\<qi^\perp\>\eta_{ia}\big)\Bigr]
    \Bigl[\delta^{(4)}_{34}\big(|i^\perp\>\eta_{ia}\big)+K_n\delta^{(2)}_{34}\big(\<qi^\perp\>\eta_{ia}\big)\Bigr]\,.
\end{equation}
This supervertex, just as the massless MHV vertex, preserves all $\tilde{Q}$ supersymmetries. Indeed,
\begin{equation}
    |\tQ_a\>{\cal V}_n~=~0\,,\qquad \text{with }~~|\tQ_a\>\equiv\sum_{i}|\tQ_{ia}\>\,.
\end{equation}
In fact, ${\cal V}_n$ is nothing but the product of the $\tQ$ supercharges:
\begin{equation}\label{VnQ}
    {\cal V}_n~=~\delta^{(8)}\Bigl(|\tQ_a\>\Bigr)
    ~=~
    \frac{1}{2^4} \prod_{a=1}^4 \sum_{i,j} \<Q_{ia}\,Q_{ja}\>
    \,.
\end{equation}
In this expression, all $\eta$-derivatives need to be carried out, so that ${\cal V}_n$ is a Grassmann polynomial with kinematic coefficients, and not a Grassmann differential operator.  Note that the $\delta$-function is indeed well-defined despite the $\eta$-derivatives in the definition of the $\tQ_a$ supercharges, because these anticommute with each other:
\begin{equation}
    \bigl\{|\tQ_a\>,|\tQ_b\>\bigr\}~=~0\,.
\end{equation}
This follows from the SUSY algebra
\begin{equation}
    \bigl\{|\tQ_{ia}\>^{\dot\alpha},|\tQ_{jb}\>^{\dot\beta}\bigr\}~=~
    -m_i\,
    \delta_{ij}
    \epsilon^{\dot\alpha\dot \beta}\,\Omega_{ab}\,,
\end{equation}
together with $\sum_im_i=0$. Crucially, it does not rely on the lines $i$ being on-shell, and thus also holds for CSW vertices with off-shell internal lines, whose angle spinors are given by the CSW prescription~(\ref{CSW}).

It is easy to verify explicitly that~(\ref{VnQ}) is equivalent to~(\ref{Vn}). For example,
\begin{equation}
\begin{split}
      \delta^{(2)}(\tQ_1)\delta^{(2)}(\tQ_2)~&=~
      \delta^{(2)}\biggl(|i^\perp\>\eta_{i1}-\frac{m_i|q\>}{\<q i^\perp \>}\frac{\partial}{\partial\eta_{i2}}\biggr)
      \delta^{(2)}\bigl(|j^\perp\>\eta_{j2}\bigr)\\
      ~&=~\frac{\delta^{(2)}_{12}\bigl(\<qi^\perp\>\eta_{ia}\bigr)}{\<qX\>^2}
      \biggl(\<Xi^\perp\>\eta_{i1}-\frac{m_i\<Xq\>}{\<q i^\perp \>}\frac{\partial}{\partial\eta_{i2}}\biggr)
      \Bigl(\<Xj^\perp\>\eta_{j2}\Bigr)\\[1ex]
      ~&=~\delta_{12}^{(4)}\bigl(|i^\perp\>\eta_{ia}\bigr)
      +K_n\,\delta_{12}^{(2)}\bigl(\<qi^\perp\>\eta_{ia}\bigr)\,,
\end{split}
\end{equation}
and similarly in the other $SU(2)$ sector.

Expressed in terms of supercharges, the massive CSW supervertex thus takes the same form as in the massless case,
\begin{equation}\label{dQcyc}
\boxed{~~
    \parbox[c]{1.5cm}{\includegraphics[width=1.5cm]{cswnptsupergen}}~\equiv~
    \frac{\delta^{(8)}\Bigl(|\tQ_a\>\Bigr)}{\<1^\perp2^\perp\>\cdots\<n^\perp1^\perp\>}\,.~~~
    }
\end{equation}
This form of the supervertex turns out to be convenient for our loop-level derivation of the CSW rules.

The form~(\ref{dQcyc}) of the supervertex also allows a convenient new representation of products of supervertices. For example, one can rewrite subdiagrams with two vertices that are connected by $M$ internal lines $P_I$ in the following way:
\begin{equation}
\begin{split}
\label{id}
    \parbox[c]{3cm}{\includegraphics[width=3cm]{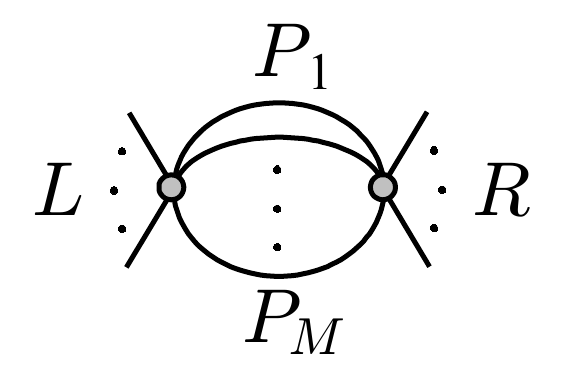}}
    ~&\propto~ \int d^{4M}\eta_{P_Ia}\,{\cal V}_{L,P_I}\,{\cal V}_{R,\dash P_I}\\[-3ex]
    ~&=~
    \int d^{4M}\eta_{P_Ia}\,\biggl[\delta^{(8)}\biggl(\,\sum_{i\in L,P_I}\!\!|\tQ_{ia}\>\biggr)\biggr]\biggl[\delta^{(8)}\biggl(\,\sum_{i\in R,\dash P_I}\!\!|\tQ_{ia}\>\biggr)\biggr]\\[1ex]
    ~&=~\delta^{(8)}\biggl(\,\sum_{i\in L,R}\!\!|\tQ_{ia}\>\biggr)\,\int d^{4M}\eta_{P_Ia}\,\delta^{(8)}\biggl(\,\sum_{i\in R,\dash P_I}\!\!|\tQ_{ia}\>\biggr)\,.
\end{split}
\end{equation}
 In the sums, we denoted the lines on the left and right vertex that do not directly connect the two vertices by $L$ and $R$, respectively.
In the final expression, the $\eta$-differentiations act throughout the expression; for example the differentiations in the left $\delta$-function can also act on the right $\delta$-function.
 This is to be contrasted with the definition of ${\cal V}_n$,~(\ref{VnQ}), where all differentiations are carried out {\em within} the $\delta$-function. In particular, the overall $\delta$-function in~(\ref{id}) {\em cannot} be simply replaced by a factor of ${\cal V}_{L,R}$!
 While trivial in the massless case, the identity~(\ref{id}) on the Coulomb branch takes some tedious but straight-forward algebra to derive. Iterating this identity, it is clear that we can pull an ``overall'' $\delta$-function out of any diagram. Specifically, any diagram with
$M$ internal lines $P_I$ can be brought into the form
\begin{equation}
    \parbox[c]{2cm}{\includegraphics[width=2cm]{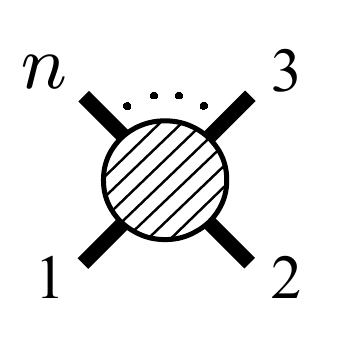}}
    ~=~\delta^{(8)}\biggl(\sum_{i=1}^n|\tQ_{ia}\>\biggr)\int d^{4M}\eta_{P_I}\bigl[\ldots\bigr]\,.
\end{equation}
On the left-hand side, the {\em striped  blob denotes any individual supervertex diagram} that contributes to the $n$-point amplitude.
Again, we do not need to impose on-shell conditions on the external lines $i$, and all differentiations in the $|\tQ_{ia}\>$ act also on the remaining $\eta$-dependence in the `$\ldots$'. This representation makes the $\tQ$-supersymmetry of each individual massive CSW diagram completely manifest.

\subsection{Manifest $Q$-supersymmetry of massive CSW rules}\label{secQ}
The massive CSW rules do not only  preserve all $\tQ$ supersymmetries, they also preserve half of the $Q$ supersymmetries, diagram by diagram, even off-shell. Indeed, by momentum conservation each vertex of each diagram is invariant under the collective shift of all $\eta$-variables,
\begin{equation}
    \eta_{ia}~\to~\eta_{ia}+ [q i^\perp ]\delta\eta_{a}\,,
\end{equation}
which is generated by the supercharge
\begin{equation}
    [q\,Q^{a}]~=~\sum_i[qi^\perp]\frac{\partial}{\partial\eta_{ia}}\,.
\end{equation}
One consequence of manifest $[q\,Q^{a}]$ supersymmetry is that all-minus amplitudes and integrands {\em vanish  diagram by diagram, even off shell}:\footnote{
Again, the blob on the left-hand side denotes any individual diagram that contributes to the all-minus amplitude.}
\begin{equation}\label{allminusvanishes}
    \parbox[c]{2cm}{\includegraphics[width=2cm]{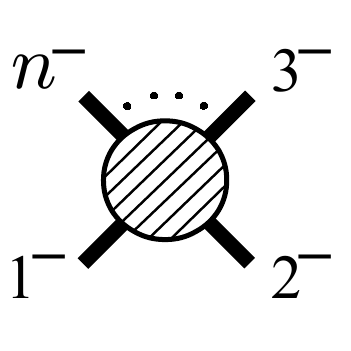}}~=~0\,.
\end{equation}
This is not surprising, because $[q\,Q^{a}]$ Ward identities can be used to show that massive all-minus amplitudes vanish~\cite{Schwinn:2006ca}. Technically, the reason for~(\ref{allminusvanishes}) is that a full fermionic integral over an integrand with a fermionic ``zero mode'' vanishes. Indeed,
\begin{equation}
\begin{split}
    \parbox[c]{2cm}{\includegraphics[width=2cm]{allminus}}
    ~&=~\int d^4\eta_{1a}d^4\eta_{2a}\cdots d^4\eta_{na}\,\parbox[c]{2cm}{\includegraphics[width=2cm]{allminusdiag}}\\[-3ex]
    ~&=~\frac{1}{[q1^\perp]^4}\int d^4\eta_{2a}\cdots d^4\eta_{na}\,\prod_b [q\,Q^{b}]
    \parbox[c]{2cm}{\includegraphics[width=2cm]{allminusdiag}}\\[-1ex]
    ~&=~0\,.
\end{split}
\end{equation}
The manipulation from the 1st line to the 2nd line
 is valid because $\eta$-integrals of all lines $i=1,\dots,n$ are present.

\subsection{Derivation of the expansion: tree amplitudes}\label{secder}

To derive the massive CSW expansion for Coulomb-branch amplitudes, we make use of on-shell recursion relations.
On-shell recursion relations are based on a complex shift $p_i\to\hat p_i(z)$ of the external momenta of an on-shell amplitude. The complex shift must preserve the on-shell condition, $\hat p_i^2=m_i^2$, and momentum conservation, $\sum_i\hat p_i=0$. Analytic properties of the $z$-dependent on-shell amplitude $\hat A_n(z)$ can then be used to express the amplitude in terms of its factorization channels, which involve products of two lower-point on-shell amplitudes. On-shell recursion relations are
straight-forwardly applicable
 when the amplitude vanishes at large $z$, $\hat A_n(z)\to0$. In that case factorization channels completely determine the amplitude. If $\hat A_n(z)$ does not vanish as $z\to\infty$, there is an additional ``boundary'' contribution
 to the recursion relation from the residue picked up by a contour around $z=\infty$,
\begin{equation}
    B_n~\equiv~\oint_{C_{\infty}}\frac{dz}{2\pi i z}
    \,
    \hat A_n(z)\,.
\end{equation}
 In particular, {\em if $\hat A_n(z)$ goes to a  constant in the large-$z$ limit}, this constant is precisely the boundary term $B_n$:
\begin{equation}
    B_n~=~\lim_{z\to\infty} \hat A_n(z)\,.
\end{equation}
The boundary term $B_n$ must be derived separately by an independent method.\footnote{
 See~\cite{Feng:2009ei,Feng:2010ku,Feng:2011tw} for a different approach to determining boundary contributions.} 
This will be of central importance for us in the following.

For massless $\cn=4$ SYM,  an anti-holomorphic all-line shift,
\begin{equation}\label{holom0}
     |\hat i]=|i]+z b_i|q]\,,\qquad \sum_i b_i|i\>=0\,,
\end{equation}
 was used in~\cite{Elvang:2008vz} to derive the tree-level CSW expansion. Here, the $b_i$ are complex numbers subject to the momentum conservation constraint given in~(\ref{holom0}), while $|q]$ is the CSW reference spinor.
Under this shift, massless  N$^k$MHV amplitudes vanish as $1/z^k$. MHV amplitudes  ($k=0$), are invariant under this shift, as they only depend on holomorphic spinors; in the language above, the massless $k=0$ amplitude has a boundary term at $z\to\infty$. In this case, the boundary term is actually the entire $k=0$ amplitude, because no additional $z$-suppressed terms appear in the Parke-Taylor expression.
In the CSW expansion, MHV amplitudes are thus supplied separately as an input into the recursion relation and constitute the basic vertices of the expansion. The diagrams in the expansion are simply all diagrams with MHV vertices, connected by scalar propagators.

More generally, all-line shift recursions relations imply a
CSW-like expansion for the amplitudes of a theory {\em if
all amplitudes that do not vanish at large $z$ have holomorphic boundary terms}:
  \begin{equation}\label{cond2}
    B_n~=~\text{holomorphic in angle spinors}\,.
  \end{equation}
In that case, any amplitude in the theory can be expressed as the sum over all diagrams with scalar propagators and boundary terms  $B_n$ as vertices. This can be proven inductively.
As always, the CSW prescription is understood for the angle spinors of internal lines in the holomorphic
 vertices.

A massive generalization of the anti-holomorphic all-line shift was introduced in~\cite{Cohen:2010mi}. It acts simply as
\begin{equation}\label{antiholo}
     |\hat i^\perp]~=~|i^\perp]+z b_i|q]\,,\qquad \sum_i b_i|i^\perp\>=0\,,
\end{equation}
and satisfies all requirements for an on-shell deformation. Crucially, the reference spinor $|q]$ here must coincide with the reference spinor used in the massive spinor helicity formalism to define massive polarization vectors. It was shown in~\cite{Cohen:2010mi} that  the large-$z$ behavior of a general amplitude in a general ($4$-dimensional) theory is given by
\begin{equation}
    \hat A_n(z)~\sim~ z^{s}\quad \text{ (or better)}\,,\quad\text{ with } 2s=4-n-c+\sum_ih_i\,.
\end{equation}
Here, $h_i$ is the $q$-helicity of the particle $i$
 (as defined after~(\ref{pol})),
and $c$ is the mass dimension of the product of couplings\footnote{If more than one product of couplings appears, $c$ is the smallest mass dimension.} that contribute to the amplitude $A_n$.

Let us apply this to an N$^{k}$MHV$\times$N$^{k'}$MHV amplitude on the Coulomb branch of $\cn=4$ SYM. We remind the reader that $k$ and $k'$ are related to the $\eta$ degrees of the superamplitude with respect to the two $SU(2)$ subsectors of the $R$-symmetry. Specifically, the N$^{k}$MHV$\times$N$^{k'}$MHV amplitude is of degree $\eta_{1,2}^{2(k+2)}\eta_{3,4}^{2(k'+2)}$.   For example, MHV amplitudes (or, more precisely, MHV$\times$MHV amplitudes) correspond to $k=k'=0$. UHV amplitudes correspond to $k=k'=-1$, while UHV$\times$MHV correspond to $k=-1$, $k'=0$, and so on.
It is easy to see that
\begin{equation}\label{falloff}
    4-n+\sum_ih_i= -(k+k')\quad \Longrightarrow \quad
    \hat A_n(z)~\sim~\frac{1}{z^{(k+k'+c)/2}}\quad\text{(or better)\,,~~~~as}~~~z \to \infty\,.
\end{equation}
The coupling dimension $c$ takes a little more thought. $SU(4)$ violation on the Coulomb branch is induced by the scalar vev. Each insertion of the scalar vev corresponds to one power of the mass $m$. $SU(4)$-violating amplitudes, for which $k\neq k'$, thus necessarily involve massive couplings. More precisely, the couplings contributing to any Coulomb-branch amplitude satisfy $c\geq |k-k'|$.  Also, we have the obvious bound $k,k'\geq -1$ because the UHV sector is the lowest
 non-vanishing
sector in the theory. It follows that $\hat A_n(z)$ vanishes at large $z$ for any amplitude with $k>0$ or $k'>0$ (or both).
The only amplitudes for which $\hat A_n(z)$ is {\em not} guaranteed to fall off at large $z$ are
\begin{equation}
\begin{split}\label{bndryamps}
   \text{MHV amplitude:}~~~c=0&\qquad\Rightarrow\qquad \hat A_n^{\rm  MHV}(z)~\sim~z^0\,,\\
   \text{UHV$\times$MHV amplitude:}~~~c=1&\qquad\Rightarrow\qquad \hat A_n^{\rm UHV\times MHV}(z)~\sim~z^0\,,\\
   \text{UHV amplitude:}~~~c=2&\qquad\Rightarrow\qquad \hat A_n^{\rm  UHV}(z)~\sim~z^0\,.
\end{split}
\end{equation}
The $c=2$ for the UHV amplitude may seem surprising as this amplitude does not violate $SU(4)$; however, it is well-known that UHV amplitudes are forbidden in the massless limit by the SUSY Ward identities. While these amplitudes are non-vanishing in the massive theory, they are suppressed by $m^2$.

The amplitudes in~(\ref{bndryamps}) are the only amplitudes with potential boundary terms at infinity. These boundary terms will be computed below, with the result:
\begin{equation}
\begin{split}\label{bndryterms}
   B_n^{\rm  MHV} ~&=~ \frac{\delta^{(8)}\big(|i^\perp\>\eta_{ia}\big)}{\<1^\perp2^\perp\>\cdots\<n^\perp1^\perp\>}\,,\\
   B_n^{\rm UHV\times MHV}~&=~ K_n\,\frac{\delta^{(4)}_{12}\big(|i^\perp\>\eta_{ia}\big)\,\delta^{(2)}_{34}\big(\<qi^\perp\>\eta_{ia}\big)
    +\delta^{(4)}_{34}\big(|i^\perp\>\eta_{ia}\big)\,\delta^{(2)}_{12}\big(\<qi^\perp\>\eta_{ia}\big)
    }{\<1^\perp2^\perp\>\cdots\<n^\perp1^\perp\>}\,,\\
  B_n^{\rm  UHV} ~&=~K_n^2\,\frac{\delta^{(4)}\big(\<qi^\perp \>\eta_{ia}\big)}{\<1^\perp2^\perp\>\cdots\<n^\perp1^\perp\>}\,.
\end{split}
\end{equation}
In particular, these boundary terms are holomorphic in angle spinors and thus satisfy~(\ref{cond2}). The criteria for the validity of a  CSW-like expansion are thus fulfilled.
Amplitudes on the Coulomb-branch of $\cn=4$ SYM can therefore be computed from a vertex expansion with scalar propagators and vertices given by the boundary terms in~(\ref{bndryterms}). This is precisely the massive CSW expansion proposed in~\cite{Kiermaier:2011cr}, whose
diagrammatic
rules we reviewed above.

Before we turn to a derivation of the boundary terms~(\ref{bndryterms}), let us state an immediate consequence of this result:
since
this massive CSW expansion is valid for any $q$, it is in particular valid for the special $q$ presented in~(\ref{spq}). It was shown in~\cite{Kiermaier:2011cr} that the soft-limit proposal~(\ref{proposal}) is equivalent to the massive CSW expansion for this special $q$. Our derivation of the massive CSW expansion from recursion relations thus provides a rigorous (albeit indirect) proof of the proposal~(\ref{proposal}),
 and its generalization to amplitudes with arbitrarily many massive lines.

\vspace{2mm}
\noindent {\bf Derivation of the boundary terms}\\
We now show that the boundary terms are given by the expressions in~(\ref{bndryterms}).
We
need to compute the amplitudes $A_n^{\rm UHV\times UHV}$, $A_n^{\rm UHV\times MHV}$, and $A_n^{\rm MHV\times MHV}$ in the limit $z\to\infty$ under the {\em anti-holomorphic} all-line shift~(\ref{antiholo}). For $n=3,4$ these boundary terms can be verified straight-forwardly from known explicit Coulomb-branch superamplitudes~\cite{Craig:2011ws}. For $n>4$,  we compute these boundary terms recursively from the parity-conjugate recursion relation:
a
{\em holomorphic} all-line shift,
\begin{equation}\label{holo}
|\tilde i^\perp\>~=~|i^\perp\>+ w\,\tilde{b}_i |q\>\,.
\end{equation}
Momentum conservation for the doubly-shifted momenta implies the following conditions on the complex parameters $\tilde{b}_i$:
\begin{equation}
    \sum \tilde{\hat p}_i~=~0
~~~~~\Rightarrow~~~~~
\sum_i\tilde{b}_i|i]~=~0\,,
\qquad \sum_ib_i\tilde{b}_i~=~0\,.
\end{equation}
Let us begin with the lowest-order boundary term, $B_n^{\rm UHV}$. We thus study $\hat A_n^{\rm UHV}(z)$ as $z\to \infty$.   Only its leading $c=2$ contribution can give rise to a non-vanishing boundary term at $z\to\infty$. Indeed, we can immediately drop all $c>2$ contributions to the amplitude, because they must vanish at large $z$ by~(\ref{falloff}). Expanding $\hat A_n^{\rm UHV}(z)$ under the holomorphic all-line-shift recursion relation~(\ref{holo}), it takes the schematic form
\begin{equation}
    \hat A_n^{\rm UHV}(z)~=~ \sum_I\biggl[\tilde{\hat A}_{n_L}^{\rm UHV}\times\frac{1}{\hat P_I^2+m_I^2}\times\tilde{\hat A}_{n_R}^{\rm UHV}\biggr]_{w=w_I}\,.
\end{equation}
Here, the sum over over $I$ denotes the sum over all factorization channels that contribute to the recursion relation.
Generically, both UHV subamplitudes are $O(m^2)$, and therefore the generic terms in the sum over $I$ are $c=4$ contributions to $\hat A^{\rm UHV}(z)$. These  vanish as $z\to\infty$ and can thus be dropped. The only exception is when one of the subamplitudes is 3-point, say $n_R=3$. 3-point on-shell UHV amplitudes are $O(1)$; they are simply the massive generalization of 3-point anti-MHV amplitudes. We conclude that
\begin{equation}
\begin{split}\label{AnUHVz}
    \hat A_n^{\rm UHV}(z)~=&~ \biggl[\int d^4\eta_{P}\,\tilde{\hat A}_{n-1}^{\rm UHV}\Bigl(\tilde{\hat P}_{12},\tilde{\hat3},\ldots,\tilde{\hat n}\Bigr)\times\frac{1}{\hat P_{12}^2+m_{12}^2}\times\tilde{\hat A}_{3}^{\rm anti-MHV}\Bigl(\tilde{\hat1},\tilde{\hat2},-\tilde{\hat P}_{12}\Bigr)\biggr]_{w=w_{12}}\\
    &~~+~\text{cyclic}~+~O(1/z)\,.
\end{split}
\end{equation}
To leading order in $1/z$, only the $c=2$ contribution to the left subamplitude  $\tilde{\hat A}^{\rm UHV}_{n-1}$ contributes. It is given by the lower-point boundary term $B^{\rm UHV}_{n-1}$, which is the input of our inductive derivation.  The massive anti-MHV 3-point amplitude, on the other hand, is simply given by
\begin{equation}
    A_{3}^{\rm anti-MHV}(1,2,3)~=~\frac{\delta^{(4)}\bigl([1^\perp2^\perp]\eta_{3a}+\text{cycl}\bigr)}{[1^\perp2^\perp][2^\perp3^\perp][3^\perp1^\perp]}\,.
\end{equation}
Let us first carry out the $\eta$-integration in~(\ref{AnUHVz}).
\begin{equation}
\begin{split}\label{etaPint}
    &\int d^4\eta_{Pa}\,\delta^{(4)}\Bigl(\<q\tilde{\hat P}^\perp_{12}\>\eta_{Pa}+\sum_{i=3}^n\<qi^\perp \>\eta_{ia}\Bigr)\times
\delta^{(4)}\Bigl([\hat 1^\perp\hat 2^\perp]\eta_{Pa}+[\hat 2^\perp\tilde{\hat P}^\perp_{12}]\eta_{1a}+[\tilde{\hat P}^\perp_{12}\hat 1^\perp]\eta_{2a}\Bigr)\\[1ex]
&=~[\hat 1^\perp\hat 2^\perp]^4\delta^{(4)}\Bigl(\,\sum_{i=1}^n\<qi^\perp \>\eta_{ia}\Bigr)\,.
\end{split}
\end{equation}
To see this, one simply uses the second $\delta$-function to eliminate the $\eta_P$-dependence in the first one, and then carries out the integration.
Next, consider the kinematic factor $\tilde{\hat K}_{n-1}$. At large $z$, we have $|\tilde{\hat P}_{12}^\perp\>\propto |\tilde{1}^\perp\>\propto |\tilde{2}^\perp\>$\,. To see this, note that at large $z$ we can neglect masses in the $3$-point anti-MHV vertices. Therefore, the angle-spinors of all lines must become proportional to each other in this limit, just as in the massless case.  We can then rewrite $\tilde{\hat K}_{n-1}$ as
\begin{equation}
\begin{split}
    \tilde{\hat K}_{n-1}
    ~&=~\frac{(m_1+m_2)\<X \tilde{\hat P}_{12}^\perp\>}{\<X q\>\<\tilde{\hat P}_{12}^\perp q\>}+\sum_{i=3}^n\frac{m_i\<X \tilde{i}^\perp\>}{\<X q\>\<i^\perp q\>}
    ~=~\sum_{i=1}^n\frac{m_i\<X \tilde{i}^\perp\>}{\<X q\>\<i^\perp q\>}+O(1/z)~=~\tilde{K}_{n}+O(1/z)\,.
\end{split}
\end{equation}
Straight-forward spinor gymnastics in the large-$z$ limit then gives
\begin{equation}
  \hat A_n^{\rm UHV}(z)~=~\Biggl[\tilde{K}_{n}^2\,\frac{ \delta^{(4)}(\<qi^\perp \>\eta_{ia})}{\<1^\perp2^\perp\>\<\tilde{2}^\perp\tilde{3}^\perp\>\cdots \<\tilde{n}^\perp\tilde{1}^\perp\>}\Biggr]_{w=w_{12}}+~\text{cyclic}+O(1/z)\,.
\end{equation}
Using Cauchy's theorem in $w$, this implies
\begin{equation}
  \hat A_n^{\rm UHV}(z)~=~{K}_{n}^2\,\frac{ \delta^{(4)}(\<qi^\perp \>\eta_{ia})}{\<1^\perp2^\perp\>\<{2}^\perp{3}^\perp\>\cdots \<{n}^\perp{1}^\perp\>}+O(1/z)
  ~=~
  B_n^{\rm UHV} +O(1/z)
  \,.
\end{equation}
In the large-$z$ limit we thus precisely recover the boundary term $B_n^{\rm UHV}$ in~(\ref{bndryterms}), which completes its derivation.

The derivation of the remaining boundary terms in~(\ref{bndryterms}) is analogous. The only step that requires a slight modification is the treatment of the $\eta_P$ integral,~(\ref{etaPint}). For example, in the computation of the boundary term $B_n^{\rm MHV}$ we carry out this integration as
\begin{equation}
\begin{split}\label{etaPint2}
    &\int d^4\eta_{Pa}\delta^{(8)}\Bigl(|\tilde{\hat P}^\perp_{12}\>\eta_{Pa}+\sum_{i=3}^n|\tilde{i}^\perp \>\eta_{ia}\Bigr)\times
\delta^{(4)}\Bigl([\hat 1^\perp\hat 2^\perp]\eta_{Pa}+[\hat 2^\perp\tilde{\hat P}^\perp_{12}]\eta_{1a}+[\tilde{\hat P}^\perp_{12}\hat 1^\perp]\eta_{2a}\Bigr)\\[1ex]
&=~[\hat 1^\perp\hat 2^\perp]^4\delta^{(8)}\Bigl(\,\sum_{i=1}^n|\tilde{i}^\perp \>\eta_{ia}\Bigr)+O(1/z)\,,
\end{split}
\end{equation}
 and similarly for $B_n^{\rm UHV\times MHV}$.

This completes our derivation of the boundary terms~(\ref{bndryterms}) and proves the massive CSW expansion proposed in~\cite{Kiermaier:2011cr} at tree level.

\subsection{Derivation of the expansion: loop integrands}

We now extend the derivation of the massive CSW expansion to  loop integrands
\footnote{To have a well-defined meaning of loop integrand, we need to assume planarity to avoid ambiguity in the labeling of loop momenta. This is the only sense in which our analysis  requires planarity.}
on the Coulomb-branch of $\cn=4$ SYM.
 We again use an antiholomorphic all-line shift recursion relation, which was previously used for massless loop integrands in~\cite{Bullimore:2010dz}. In addition to the external momenta $p_i$, which are shifted as in~(\ref{antiholo}), we also need to shift the $L$ independent loop momenta $\ell^{(1)},\ldots,\ell^{(L)}$. We choose
\begin{equation}
    \ell^{(i)}~\to~\hat \ell^{(i)}~=~\ell^{(i)}+z\, b^{(i)}\,q\,,
\end{equation}
where the  $b^{(i)}$, $i=1,\ldots,L$ are arbitrary complex numbers. This is the momentum-space analog of the loop-integrand all-line shift in twistor space introduced in~\cite{Bullimore:2010dz}. At large $z$, the N$^{k}$MHV$\times$N$^{k'}$MHV integrand goes as
\begin{equation}\label{loopfalloff}
    \hat{\cal I}^L_n~\sim~\frac{1}{z^{(k+k'+c)/2+2L}}\quad\text{(or better)\,,~~~~as}~~~z \to \infty\,.
\end{equation}
This follows immediately from dimensional analysis and the little-group properties. As at tree level, we have the additional constraints $k,k'\geq  -1$ and $c\geq0$. It follows that any integrand vanishes at large $z$:
\begin{equation}\label{loopz}
    \hat{\cal I}^L_n~\to~0\qquad\text{ as }~z\to\infty\,.
\end{equation}
Unlike at tree level,
 there are thus no ``boundary terms'' in the recursion relation at loop-level.

We remind the reader of the general structure of recursion relations for an $n$-point $L$-loop integrand $\hat {\cal I}_{n}^{L}$~\cite{ArkaniHamed:2010kv,Boels:2010nw}. The recursion relation contains ``conventional'' factorization channels into two on-shell subintegrands $\hat {\cal I}^{L_1}_{n_1}$ and $\hat {\cal I}_{n_2}^{L_2}$ satisfying $n_1+n_2=n+2$ and $L_1+L_2=L$:
\begin{equation}
    \sum_{I}
    ~~\parbox[c]{3.2cm}{\includegraphics[height=1.5cm]{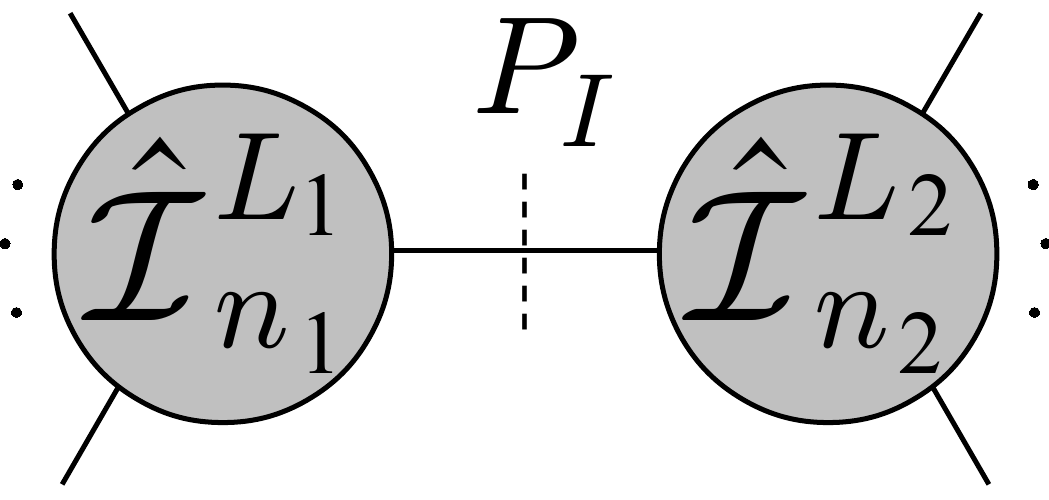}}\,.
\end{equation}
Factorization channels with ``zero-loop'' integrands, which are  just tree-level amplitudes  (${\cal I}^0_{n_i}=A_{n_i}^{\rm tree}$), are of course included in this sum. In addition to these conventional factorization channels, there are also factorizations that are not ``1-particle reducible''; in that case, the cut loop propagator $\hat P_I$ appears as an incoming
{\em and} outgoing on-shell momentum in a single subintegrand; these contributions thus contain $(L\!-\!1)$-loop subintegrands that are evaluated in the forward limit, $\hat {\cal I}^{L-1}_{n+2}(\hat P_I,\dash\hat P_I,\hat 1,\ldots,\hat n)$:
\begin{equation}
    \sum_{I}~~\parbox[c]{3.2cm}{\includegraphics[height=1.5cm]{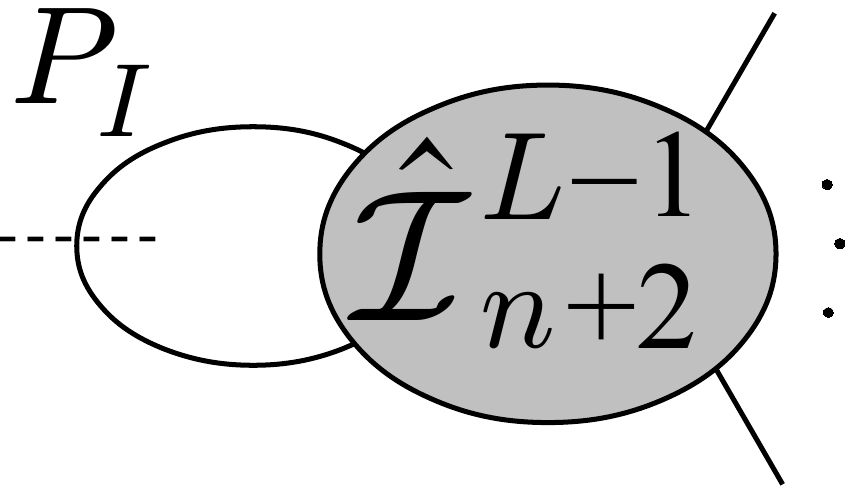}}\,.
\end{equation}
 In summary, loop-level recursion relations take the schematic form~\cite{ArkaniHamed:2010kv,Boels:2010nw}
\begin{equation}\label{looprec}
     {\cal I}_{n}^{L}~~=~\sum_{\text{reducible }I}
    ~~\parbox[c]{3.2cm}{\includegraphics[height=1.5cm]{cswloop1PR}}~+~\sum_{\text{irreducible }I}~~\parbox[c]{3.2cm}{\includegraphics[height=1.5cm]{cswloop1PI}}
    \,.
\end{equation}

At tree level, the large-$z$ falloff of the amplitude is sufficient for the validity of recursion relations. At loop level, however, one encounters a new condition: the forward integrands on the right-hand side of~(\ref{looprec}) must be well-defined. It was argued in~\cite{CaronHuot:2010zt} that $\cn=2$ SUSY is sufficient for well-defined forward limits of massive 1-loop amplitudes.
At any loop order, the potential subtleties of forward limits arise from contributions with self-energy-type subdiagrams; such subdiagrams contain the same propagator \emph{twice}. This makes factorization of such diagrams a subtle issue.

For $\cn=4$ SYM on the Coulomb bramch, we expect the forward limit of  integrands to be well-defined. Indeed,  self-energy-type diagrams should vanish as as the momentum $P$ is taken on-shell. This zero cancels the additional propagator in the denominator. However, as we will show now, a much stronger statement holds when the massive CSW expansion is used. Indeed, \emph{all self-energy-type contributions vanish diagram-by-diagram, at any loop order, even when the momentum $P$ is off-shell}:
\begin{equation}\label{SEdiag}
    \parbox[c]{2cm}{\includegraphics[width=2cm]{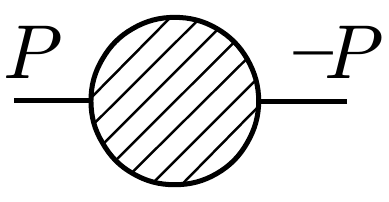}}~~=~0\,.
\end{equation}
In particular, all diagrams which contain a self-energy-type subdiagram vanish.

To see this, it is instructive to first consider the one-loop level. The only self-energy-type diagram in this case is the bubble diagram,
which we write in terms of the supervertex \reef{superV} as
\begin{equation}
\begin{split}\label{bub}
    \parbox[c]{2cm}{\includegraphics[width=2cm]{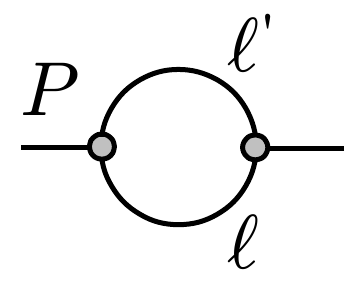}}
    ~&=~
    \parbox[c]{2cm}{\includegraphics[width=2cm]{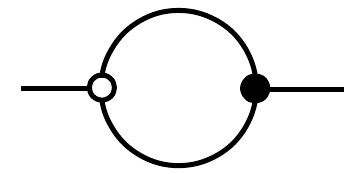}}
    +\parbox[c]{2cm}{\includegraphics[width=2cm]{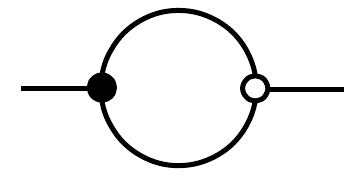}}
    +\parbox[c]{2cm}{\includegraphics[width=2cm]{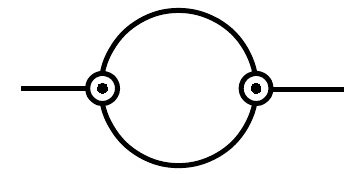}}\\
    ~&=~
    \frac{\<\ell\ell'\>^4\delta^{(4)}\bigl(\<qP\>\eta_{P,a}\!-\!\<qP\>\eta_{\dash P,a}\bigr)}{(\<P\ell\>\<\ell\ell'\>\<\ell'P\>)^2}\times\bigl[
    K_L^2+K_R^2+2K_LK_R\bigr]\,.
\end{split}
\end{equation}
with all other combinations of component vertices vanishing.
The kinematic factors associated with the left and right vertices are given by
\begin{equation}\label{KLKR}
    K_L=m_\ell\frac{\<P\ell\>}{\<Pq\>\<\ell q\>}
    +m_{\ell'}\frac{\<P\ell'\>}{\<Pq\>\<\ell' q\>}\,, \qquad
    K_R=-m_{\ell}\frac{\<P\ell\>}{\<Pq\>\<\ell q\>}
    -m_{\ell'}\frac{\<P\ell'\>}{\<Pq\>\<\ell' q\>} \,.
\end{equation}
 Clearly, $K_L+K_R=0$, and the bubble diagram~(\ref{bub}) vanishes,
\begin{equation}\label{oneloopbub}
\parbox[c]{2cm}{\includegraphics[width=2cm]{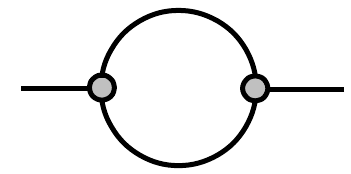}}~=~0
\,.
\end{equation}
At two loops, explicit computation also verifies the vanishing,
before integration,
of
each of
the three non-trivial self-energy topologies
\begin{equation}
    \parbox[c]{2cm}{\includegraphics[width=2cm]{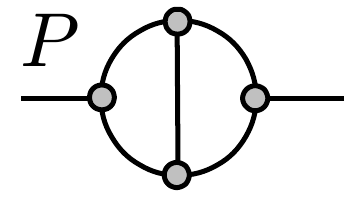}}
    ~=~\parbox[c]{2cm}{\includegraphics[width=2cm]{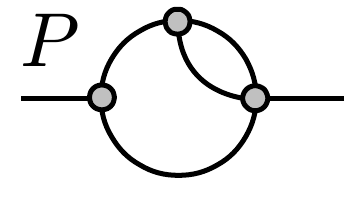}}
    ~=~\parbox[c]{2cm}{\includegraphics[width=2cm]{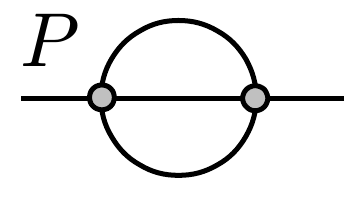}}
    ~=~0
    \,.
\end{equation}
This again holds for generic \emph{off-shell} momentum $P$.
All other two-loop self-energy diagrams contain the subdiagram~(\ref{oneloopbub}), and thus vanish by the  one-loop computation.

The fact that self-energy CSW diagrams vanish individually even off-shell at one and two loops is not a coincidence; in fact this holds at any loop order as a consequence of the manifest  supersymmetries of the massive CSW rules discussed in sections~\ref{sectQ} and~\ref{secQ}. To see this, consider an arbitrary $L$-loop self-energy diagram. For simplicity, we combine all propagators and cyclic spinor brackets of the vertices into one finite overall kinematic constant $C$ and focus on the $\eta$-dependence,
\begin{equation}
    \parbox[c]{2cm}{\includegraphics[width=2cm]{SEallloop}}~=~C\int\prod_rd^4\eta_{\ell_{r}a}\prod_v {\cal V}_v\,.
\end{equation}
Here, the labels $r$ and $v$ enumerate  internal lines and supervertices~(\ref{Vn}) of the diagram, respectively. It is convenient to single out one vertex ${\cal V}_1$ as special, say the vertex that contains the line $-P$. This vertex is connected by $M$ internal lines $\ell_1,\ldots,\ell_M$ to the remaining diagram:
\begin{equation}
    \parbox[c]{2cm}{\includegraphics[width=2cm]{SEallloop}}~~~=~~~
    \parbox[c]{2.75cm}{\includegraphics[width=2.75cm]{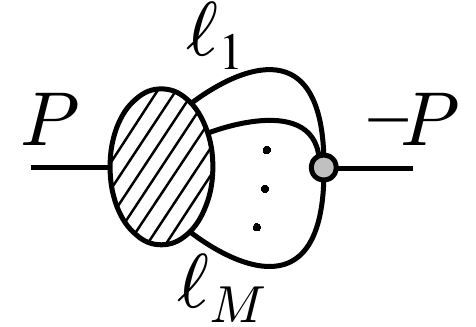}}\,.
\end{equation}
We can use the identity~(\ref{id}) iteratively to turn  the supervertex ${\cal V}_1$ into an ``overall'' $\delta^{(8)}(\tQ)$ of the external lines; for our self-energy diagram there are only two such external lines of momentum $P$ and $-P$, and a short computation shows that
\begin{equation}\label{dQ2line}
\begin{split}
    \delta^{(8)}\bigl(|\tQ_{P,a}\>+|\tQ_{\dash P,a}\>\bigr)~=&~m_P^4\,\delta^{(4)}\big(\eta_{P,a}-\eta_{\dash P,a}\big)\prod_a\Bigl(\frac{\partial}{\partial\eta_{P,a}}+\frac{\partial}{\partial\eta_{\dash P,a}}\Bigr)\\
    ~\to&~m_P^4\,\delta^{(4)}\big(\eta_{P,a}-\eta_{\dash P,a}\big)\prod_a\frac{\partial}{\partial\eta_{P,a}}\,.
\end{split}
\end{equation}
This operator acts on the remaining vertices. In the last step we used that the remaining vertices are independent of $\eta_{\dash P, a}$.
The intermediate state sums and the differentiations in~(\ref{dQ2line}) project out an ``all-minus'' diagram with lines $P,\ell_1,\ldots,\ell_M$.
Schematically,
\begin{equation}\label{toallminus}
    \parbox[c]{2.75cm}{\includegraphics[width=2.75cm]{SEallloopspecial}}~=~C\,m_P^4\,
    \delta^{(4)}\big(\eta_{P,a}-\eta_{\dash P,a}\big)\,\times\!\parbox[c]{2.75cm}{\includegraphics[width=2.75cm]{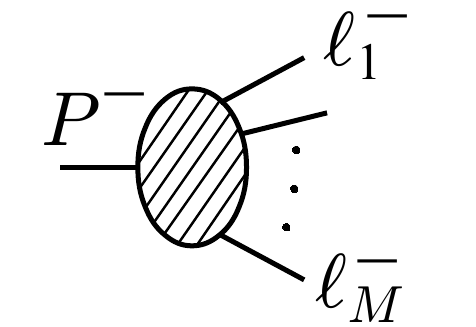}}\,.
\end{equation}
However, by~(\ref{allminusvanishes}), any all-minus diagram vanishes individually, even off-shell!  This is a direct consequence of the $[q\,Q^{a}]$ supersymmetry of massive CSW diagrams, as discussed in section~\ref{secQ}. Therefore the right-hand side of~(\ref{toallminus}) vanishes, and we conclude that~(\ref{SEdiag}) holds: massive CSW self-energy diagrams vanish individually off-shell at any loop order.

\medskip

It is now easy to prove that the loop integrand of $\cn=4$ SYM on the Coulomb branch is given by all massive CSW diagrams that do not contain any self-energy-type subdiagrams. In fact, this  follows immediately from all-line shift recursion relations: the recursion relations are valid because, by~(\ref{loopz}), any integrand vanishes at large $z$ under the  anti-holomorphic all-line shift. Also, all forward integrands entering the recursion relation are well-defined and straight-forwardly given by their CSW expression, because self-energy-type subdiagrams are manifestly absent from the expansion. This completes our proof of the massive CSW expansion for loop amplitudes on the Coulomb branch of $\cn=4$ SYM.

\section*{Acknowledgements}
We thank
R.~Akhouri,
N.~Arkani-Hamed, S.~Badger, Z.~Bern, R.~Boels, S.~Caron-Huot, J.-J.~Carrasco, T.~Cohen, L.~Dixon, H.~Johansson, and D.~Skinner for valuable discussions.
We would like to particularly thank C.~Peng for valuable work on the massive supervertex.
The research of DZF is supported by NSF grant PHY-0967299 and by
the US Department of Energy through cooperative research agreement DE-FG-0205FR41360.
HE is supported by NSF CAREER Grant PHY-0953232, and in part by the US Department of Energy under DOE grants DE-FG02-95ER 40899.
The research of MK is supported by
NSF grant PHY-0756966.

\appendix
\setcounter{equation}{0}
\section{Matching the all-plus integrand to known expressions}\label{appmatch}

Let us first study the $n=3$ case in some detail.
Expanding the first trace in~(\ref{matchID}) for $n=3$, we have
\bea
 \nonumber
\Tr_-\big[(\ell_{1} \ell_{2} + \mu^2)(\ell_{2} \ell_{3} + \mu^2)(\ell_{3} \ell_{1} + \mu^2)\big]
&=& 2 \ell_1^2 \ell_2^2 \ell_3^2
   + \mu^2 \big( \ell_1^2 \Tr_-[\ell_3\ell_2] +  \ell_2^2 \Tr_-[\ell_1\ell_3]
   + \ell_3^2 \Tr_-[\ell_2\ell_1] \big) \\[1mm]
  &&
   +
   \mu^4 \big(
   \Tr_-[\ell_3\ell_2] +  \Tr_-[\ell_1\ell_3]
   + \Tr_-[\ell_2\ell_1]
   \big)
   +2\mu^6 \,,
   \label{n3tr1}
\eea
and the second trace in~(\ref{matchID}) gives
\begin{equation}
 \Tr_-\big[ d_1 d_2 d_3 \big]
 ~=~
 2 \ell_1^2 \ell_2^2 \ell_3^2
 +  2 \mu^2 \big( \ell_1^2 \ell_2^2 +  \ell_1^2 \ell_3^2 +  \ell_2^2 \ell_3^2 \big)
   +
   2\mu^4 \big(  \ell_1^2+  \ell_2^2 +  \ell_3^2   \big)
   +2\mu^6 \, .
\label{n3tr2}
\end{equation}
Using that
\bea \label{TrHelp}
 \Tr_-(\ell_{i+1}\ell_i) = 2 \ell_{i+1}\ell_i =
- (\ell_{i+1}-\ell_i)^2+\ell_{i+1}^2+\ell_i^2 = \ell_{i+1}^2+\ell_i^2
\eea
(because
$\ell_{i+1}-\ell_i = p_{i+1}$ is null), we find that the two expressions \reef{n3tr1} and \reef{n3tr2} are identical and thus cancel to directly give~(\ref{Ippp}), $I^{+++}(1,2,3) = 0$.

Next we turn to the 4-point all-plus integrand. We again organize the expansion of the traces in powers of $\mu$. This time we need both \reef{TrHelp} as well as reductions of
traces $\Tr_- \,(\,\pslash_i\, \pslash_j \, \pslash_k\,  \pslash_l)$;
for the latter we benefit from the fact that the 4-point answer cannot contain parity-odd terms. Systematically converting dot-products of $\ell_i$ to $\ell_i^2$ using identities such as \reef{TrHelp}, we find that all terms cancel (before integration) except a $\mu^4$-term
$\Tr_- \,(\,\pslash_1\, \pslash_2 \, \pslash_3\,  \pslash_4)$.
We note that
\bea
  \frac{\Tr_-\,(\,\pslash_1\, \pslash_2 \, \pslash_3\,  \pslash_4)}{\<12\>\<23\>\<34\>\<41\>}
  ~=~
  \frac{[12][34]}{\<12\>\<34\>}
\eea
so that our final answer is
\bea
  I^{++++}(1,2,3,4) \simeq 2N_p\frac{[12][34]}{\<12\>\<34\>} \frac{\mu^4}{d_1d_2d_3d_4}\,,
\eea
as presented in~(\ref{Ipppp}).

Finally, let us treat the $n=5$ case. Consider first the parity-even contributions. These work out almost as in the $n=4$ case: we complete dot-products to $\ell_i^2$'s and leftover Mandelstam invariants of external momenta. Everything cancels except for two compact expressions in the $\mu^4$- and $\mu^6$-terms: these two expressions combine to
$-(s_{12} s_{23} \,d_4 + \text{cyclic})$.
The simplifications leading to this involve traces
$\Tr_- \,(\,\pslash_i\, \pslash_j \, \pslash_k\,  \pslash_l)$,
and this time we cannot discard the parity-odd contributions. The $\mu^6$-terms simplify directly to $2i \eps(1,2,3,4)$. The $\mu^4$-terms on the other hand are more interesting: they can be written
 as
\begin{equation}
  2i \mu^4 \big( - \ell_5^2 \eps(1,2,3,4) - (d_4-d_3) \eps(\ell123)
  - (d_3-d_2) \eps(\ell124)
  - (d_2-d_1) \eps(\ell134)
  - (d_1-d_4) \eps(\ell234)
  \big) \,.
  \label{n5podd}
\end{equation}
In the integrand this must be divided by $d_1d_2d_3d_4d_5$.
Consider first $d_4\eps(\ell123)$.
Note that the integral $\int d^D\ell~ \ell^\mu/(d_1d_2d_3d_5)$ only knows about $p_1,p_2,p_3$ and thus it can be expressed as a linear combination of those three vectors. Thus when contracted into $\eps(\ell123)$ we get zero.
Next, consider $d_3 \eps(\ell123)$ and $d_3 \eps(\ell124)$. The integrals
$\int d^D\ell~ \ell^\mu/(d_1d_2d_4d_5)$ must be a linear combination of the three vectors $p_1$, $p_2$ and $p_3+p_4$. If the coefficient of the $(p_3+p_4)$-term is $c_3$, then integration gives $d_3 \eps(\ell123)+d_3 \eps(\ell124) \to c_3 \eps(4123)+c_3 \eps(3124)  = 0$. Likewise one can show that the pair of $d_2$-terms and the pair of $d_1$-terms in \reef{n5podd} cancel after integration.
Finally, the last integral $-d_4\eps(\ell234)$ can be evaluated after a shift $\ell \to \ell - p_1$; the non-vanishing contribution comes from the $-p_1$ in the numerator and is the integral of $\eps(1234)/(d_1d_2d_3d_4)$. Taking this back to the integrand-level, we can now write \reef{n5podd}
\bea
  2i \mu^4 \big( - \ell_5^2 \eps(1,2,3,4)+ d_5\eps(1,2,3,4) \big)
  = 2i \mu^6 \eps(1,2,3,4) \,.
\eea
We have
carefully kept  track of the sign and now see that this contribution adds to the one we found from the $\mu^6$-terms in the trace-expansion. Combining parity-even and parity-odd terms we thus arrive at the  integrand presented in~(\ref{Ippppp})
\bea
\label{n5res}
  I^{+++++} ~\simeq~
  \frac{2N_p}{\<12\>\<23\>\<34\>\<45\>\<51\>}
  \bigg(- \frac{1}{2}
  \bigg[
    \frac{\mu^4\,s_{12}s_{23}}{d_1d_2d_3d_5} + \text{cyclic}
  \bigg] + \frac{4i\mu^6\, \eps(1234) }{d_1d_2d_3d_4d_5}\bigg) \, .
\eea

\small
\renewcommand{\baselinestretch}{.9}

\providecommand{\href}[2]{#2}\begingroup\raggedright\endgroup

\end{document}